\documentclass[journal]{IEEEtai}
\usepackage[colorlinks,urlcolor=blue,linkcolor=blue,citecolor=blue]{hyperref}
\usepackage{graphicx}
\usepackage{graphics}
\usepackage{caption}
\usepackage{lipsum}
\usepackage{orcidlink}
\usepackage[T1]{fontenc}
\usepackage[utf8]{inputenc}
\usepackage[british,UKenglish,USenglish,english,american]{babel}
\usepackage{courier}
\usepackage{authblk}
\usepackage{subcaption}
\usepackage[square,numbers]{natbib}
\setcitestyle{numbers}

\usepackage{mdframed}
\newmdenv[backgroundcolor=yellow, linecolor=yellow]{highlight}

\usepackage{booktabs,siunitx,array,threeparttable, makecell}
\usepackage{soul, color}
\usepackage{xcolor}
\usepackage{amssymb}
\usepackage{multirow}
\usepackage{makecell}
\usepackage{amsmath}
\usepackage{float}
\usepackage{listings}
\usepackage{alltt} 
\usepackage{natbib}
\usepackage{pifont}
\definecolor{darkgreen}{RGB}{0,100,0}

\newcommand{\cmark}{\textcolor{darkgreen}{\Large \ding{51}}} 
\newcommand{\xmark}{\textcolor{red}{\Large \ding{55}}}   

\usepackage{color,array}

\begin{document}

\title{Multi-Agent Actor-Critic Generative AI for Query Resolution and Analysis} 

\author[]{}
\author{\IEEEauthorblockN{Mohammad Wali Ur Rahman\,\orcidlink{0000-0002-5009-221X}\IEEEauthorrefmark{1}, 
Ric Nevarez\IEEEauthorrefmark{2}, Lamia Tasnim Mim\IEEEauthorrefmark{3},
Salim Hariri\,\orcidlink{0000-0003-3956-3401}\IEEEauthorrefmark{1} \textit{Senior Member, IEEE}}\\
\thanks{This work is supported by National Science Foundation (NSF) projects 1624668, 1921485 and 2335046 the Department of Energy- National Nuclear Security Administration under Award Number DE-NA0003946 and the AGILITY project 4263090, sponsored by Korea Institute for Advancement of Technology (KIAT South Korea). \\
\indent \IEEEauthorrefmark{1} Mohammad Wali Ur Rahman and Dr. Salim Hariri are with the Department of Electrical and Computer Engineering from the University of Arizona, Tucson, AZ 85721, USA. 
Email: \{mwrahman, hariri\}@arizona.edu \\
\indent \IEEEauthorrefmark{2} Ric Nevarez is with Trustweb, New York, NY 10001, USA.
Email: ric@thetrustweb.com \\
\indent \IEEEauthorrefmark{3} Lamia Tasnim Mim is a Machine Learning Engineer in Avirtek, Inc.
Email: lamiya.tasnim@avirtek.com
}
}

\markboth{Journal of IEEE Transactions on Artificial Intelligence, Vol. XX, No. X, Month 2025}
{Rahman \MakeLowercase{\textit{et al.}}: Multi-Agent Actor-Critic Generative AI for Query Resolution and Analysis}

\maketitle

\begin{abstract}
In this paper, we introduce MASQRAD (Multi-Agent Strategic Query Resolution and Diagnostic tool), a transformative framework for query resolution based on the actor-critic model, which utilizes multiple generative AI agents. MASQRAD is excellent at translating imprecise or ambiguous user inquiries into precise and actionable requests. This framework generates pertinent visualizations and responses to these focused queries, as well as thorough analyses and insightful interpretations for users. MASQRAD addresses the common shortcomings of existing solutions in domains that demand fast and precise data interpretation, such as their incapacity to successfully apply AI for generating actionable insights and their challenges with the inherent ambiguity of user queries. MASQRAD functions as a sophisticated multi-agent system but “masquerades” to users as a single AI entity, which lowers errors and enhances data interaction. This approach makes use of three primary AI agents: Actor Generative AI, Critic Generative AI, and Expert Analysis Generative AI. Each is crucial for creating, enhancing, and evaluating data interactions. The Actor AI generates Python scripts to generate data visualizations from large datasets within operational constraints, and the Critic AI rigorously refines these scripts through multi-agent debate. Finally, the Expert Analysis AI contextualizes the outcomes to aid in decision-making. With an accuracy rate of 87\% when handling tasks related to natural language visualization, MASQRAD establishes new benchmarks for automated data interpretation and showcases a noteworthy advancement that has the potential to revolutionize AI-driven applications.
\end{abstract}

\begin{IEEEImpStatement}
By converting imprecise user queries into precise, actionable insights, MASQRAD transforms data analysis and improves the efficiency and accuracy of data interpretation in a variety of industries. With the integration of a multi-agent actor-critic model, MASQRAD achieves an 87\% accuracy rate in converting natural language to visualizations while also significantly minimizing errors and optimizing data utility. This development is especially significant for industries like healthcare, finance, and public policy where quick and accurate data analysis is essential. The framework enables smooth user interaction and strong analytical performance by "masquerading" as a single AI entity and operating with complex inter-agent dynamics. In addition to enhancing decision-making, MASQRAD's creative use of generative AI agents for data creation, refinement, and analysis also establishes new benchmarks for automated data interpretation, exemplifying a transformative approach in AI applications and generating new paradigms for smart, multi-agent technology integration.
\end{IEEEImpStatement}

\begin{IEEEkeywords}
Generative AI, Actor-Critic Model, Multi-Agent System, Query Resolution, Data Visualization, Deep Learning, Natural Language Processing.
\end{IEEEkeywords}

\section{Introduction}

\IEEEPARstart{W}{ith} applications ranging from image synthesis to natural language processing and beyond, generative AI has made tremendous strides in recent years. Still, there are several obstacles that need to be overcome before generative AI can be effectively used for data analysis and complex query resolution. A significant challenge is the propensity of generative models to generate “hallucinations,” or results that appear credible but are factually false or unrelated to the user’s query \cite{brown2020language, radford2019language}. When dealing with ambiguous or poorly defined user inputs, this problem is especially noticeable and can produce results that are unreliable or deceptive. In applications where accuracy and precision are crucial, hallucinations not only compromise the reliability of AI systems but also present serious difficulties.

The scalability of generative AI systems when handling big datasets presents another difficulty. The constraints imposed by API query limits and computational resources frequently cause traditional generative AI models to falter \cite{vaswani2017attention}. This limits their capacity to effectively manage large amounts of data, which lowers the caliber and comprehensiveness of the produced outputs. Longer processing times, higher computational costs, and the inability to handle real-time or nearly real-time data processing—all of which are crucial in many industrial applications—are all consequences of scalability problems.

Furthermore, a useful mechanism for creating and validating intricate analytical workflows is frequently absent from current systems. While code snippets and simple visualizations can be produced by generative AI, significant human intervention is usually necessary to ensure these outputs’ accuracy, effectiveness, and utility  \cite{he2016deep, hochreiter1997long}. This restricts the use of generative AI in fields where accuracy and in-depth research are essential. The time-consuming and error-prone nature of the manual review and correction process further reduces the effectiveness of generative AI applications.

Further complexity is added when multi-agent systems are integrated into generative AI frameworks. Conventional single-agent systems are frequently constrained by their incapacity to oversee and optimize several tasks at once. Comparatively, multi-agent systems are able to divide up the work and work together to produce more accurate and efficient outcomes. Coordinating several agents, ensuring smooth communication, and keeping the system cohesive are still difficult tasks.

To address these challenges, we propose a novel framework for generative AI that leverages a multi-agent actor-critic model to enhance query resolution and data analysis. Our system consists of three primary generative AI agents: the Actor Generative AI, the Critic Generative AI, and the Expert Analysis Generative AI, each performing distinct roles to ensure accurate and efficient query handling.

The procedure starts with the user submitting a query, which is then refined to remove ambiguities and avoid hallucinations by a trained deep learning model \cite{kingma2014adam, lecun2015deep}. After this query has been refined, it is sent to Actor Generative AI, which provides essential indicators or clues and creates a Python script to answer the query. The system’s scalability can be improved by managing large datasets efficiently and within query limits through the strategic solution and production of executable code \cite{silver2016mastering}. By preventing the dataset from being sent through the API directly, this method guarantees that bigger datasets can be handled by the system without compromising performance.

The produced Python script is examined by the Critic Generative AI to guarantee its accuracy, efficiency, and validity. Similar to a multi-agent discussion, this review procedure entails iterative refinement to maximize the script’s effectiveness and readability \cite{karras2019style}. By confirming the accuracy of the script and refining its performance, the Critic AI reduces the possibility of mistakes and raises the standard of the final product. Ensuring that the generated scripts are syntactically correct, semantically aligned with the user’s query, and capable of yielding meaningful results is a critical responsibility of the Critic AI.

After the script has been optimized and validated, it is executed to produce dataframes, numerical responses, and visualizations. Afterward, these outputs are examined by the Expert Analysis Generative AI, which evaluates the outcomes in light of the initial query. This agent offers the user insights and contextual understanding by providing a comprehensive report that explains the answers \cite{goodfellow2014generative}. The job of the Expert Analysis AI is to analyze the data, make intelligible inferences, and produce thorough reports that are comprehensible and useful.

The main contributions of our proposed MASQRAD framework are outlined as follows:
\begin{enumerate}
    \item \textbf{Enhanced Accuracy and Relevance:} Our framework significantly improves the accuracy and relevance of query responses by employing a multi-agent approach. This method effectively distributes tasks among specialized agents, ensuring that each step of the query resolution process is handled with precision and expertise.
    \item \textbf{Comprehensive Solution for Complex Tasks:} The multifaceted strategy of our system not only enhances the reliability but also provides a comprehensive solution to complex data analysis tasks. By integrating multiple generative AI agents, MASQRAD addresses various aspects of the analysis process, ensuring thorough examination and interpretation of data.
    \item \textbf{Innovative Approach to Generative AI Challenges:} Our framework addresses the key challenges of generative AI, such as hallucination errors, working with large datasets and workflow validation. We achieve this by leveraging the strengths of deep learning, strategic multi-agent systems, and code generation techniques.
\end{enumerate}
These contributions underscore our system’s potential to transform the data visualization and analysis landscape, paving the way for more accurate, insightful, and user-friendly AI-driven tools. The remainder of this paper is organized as follows: Section II reviews related work in the field of natural language to visualization tasks, providing context for our contributions. Section III details the MASQRAD framework, elaborating on the design and functionality of its multi-agent system. Section IV describes the experimental settings and results to validate the effectiveness of our framework. Finally, Section V concludes the paper with a summary of our findings and discusses potential future research directions.

\section{Related Works}

\begin{table*}[h!]
\centering
\begin{tabular}{lccccc}
\hline \hline
\textbf{System} & \textbf{Handles Vague Queries} & \textbf{Generates Code} & \textbf{Supports Large Datasets} & \textbf{Ensures Validity} & \textbf{Provides Analysis} \\ \hline 
ELIZA \cite{weizenbaum1966eliza} & \xmark & \xmark & \xmark & \xmark & \xmark \\ 
SHRDLU \cite{winograd1971procedures} & \xmark & \xmark & \xmark & \xmark & \xmark \\ 
IBM Watson \cite{ferrucci2010building} & \cmark & \cmark & \xmark & \cmark & \xmark \\ 
BERT \cite{devlin2018bert} & \cmark & \xmark & \xmark & \xmark & \xmark \\ 
GPT-3 \cite{radford2019language} & \cmark & \cmark & \xmark & \xmark & \cmark \\ 
Chat2VIS \cite{chat2vis} & \cmark & \cmark & \cmark & \xmark & \xmark \\ 
Our System (MASQRAD) & \cmark & \cmark & \cmark & \cmark & \cmark \\ \hline \hline
\end{tabular}
\caption{Qualitative Comparison of System Capabilities Across Selected Criteria}
\label{tab:comparison}
\end{table*}

Over the past few decades, natural language processing (NLP) has undergone significant evolution, with notable breakthroughs in data-driven and symbolic approaches to query resolution systems. The groundwork for early NLP research was established by symbolic approaches, which depend on rule-based systems and organized knowledge. Conversely, the introduction of deep learning and massive data processing has caused a shift in emphasis towards data-driven approaches that utilize machine learning and statistical models. The main advancements in data-driven and symbolic natural language processing (NLP) approaches are reviewed in this section, along with their advantages and disadvantages. The transition towards combining both paradigms for more reliable and effective query resolution systems is also discussed.

\subsection{Symbolic NLP Approaches}

Rule-based systems and symbolic reasoning are used in symbolic approaches to natural language processing (NLP) in order to generate and interpret human language. These approaches, which are distinguished by their reliance on predetermined rules and organized knowledge bases, have long been a mainstay of NLP research. 

ELIZA, created by Joseph Weizenbaum in the 1960s, is one of the most renowned early systems in symbolic NLP. By employing pattern matching and substitution techniques, ELIZA was able to replicate human communication, showcasing the potential of rule-based systems to imitate social interaction \cite{weizenbaum1966eliza}. ELIZA established the foundation for more complex symbolic systems despite being simple by today's standards.

Symbolic natural language processing (NLP) was further enhanced in the 1970s and 1980s by systems such as SHRDLU and LUNAR. The potential of combining language understanding with a well-defined domain was demonstrated by Terry Winograd's SHRDLU, which could comprehend and carry out commands in a block world \cite{winograd1971procedures}. William A. developed LUNAR. Woods showed how to use a structured database and natural language interface to provide answers to inquiries regarding the geological examination of moon rocks \cite{woods1973progress}.

Machine learning techniques have recently been incorporated into symbolic natural language processing (NLP). Systems that understand and respond to complicated queries, like IBM Watson, use a combination of statistical techniques and symbolic reasoning. Watson's achievements demonstrated the efficiency of hybrid strategies that blend statistical models' flexibility with symbolic methods' accuracy \cite{ferrucci2010building}.

Despite their successes, scalability and flexibility remain major challenges for symbolic approaches. Rule-based systems are inflexible by design, making it challenging to accommodate the enormous variety and ambiguity in natural language. NLP research has consequently moved toward more statistical and data-driven methodologies.

\subsection{Data-Driven NLP Approaches}

Deep learning has completely changed natural language processing (NLP) by making it possible to create models that can learn from massive datasets directly. Because of these data-driven approaches' superior performance on various tasks, symbolic methods have been largely replaced. 

Among the most important models in this paradigm change is the Transformer architecture, which Vaswani et al. introduced. \cite{vaswani2017attention}. Unlike earlier architectures, the Transformer model processes input sequences using self-attention mechanisms, enabling it to capture contextual information and long-range dependencies better. The Transformer is now the foundation of many modern NLP models, such as GPT and BERT.

Developed by Devlin et al., BERT (Bidirectional Encoder Representations from Transformers) is a noteworthy development in pre-trained language models \cite{devlin2018bert}. BERT is incredibly useful for a range of natural language processing (NLP) tasks, including text classification \cite{rahman2022bert} and question answering \cite{qu2019bert} because it is built to comprehend a word's context in both directions. Its broad adoption in academia and industry can be attributed to its ability to be tailored for particular tasks. 

Likewise, OpenAI's GPT (Generative Pre-trained Transformer) series has proven to be exceptionally proficient in language modeling and text generation \cite{radford2019language}. With only a small amount of task-specific training, GPT-3, with its 175 billion parameters, is capable of executing tasks like creative writing and code generation. Its outstanding performance highlights how large-scale pre-trained models can be generalized across a variety of NLP applications.

Other noteworthy contributions, besides BERT and GPT, are T5 (Text-To-Text Transfer Transformer) and ELMo (Embeddings from Language Models). Peters et al. developed ELMo, by taking into account the complete sentence context, developed deep contextualized word representations that dramatically enhanced performance on a variety of NLP tasks \cite{peters2018deep}. Raffel et al. introduced T5 and showed that a single strategy can produce state-of-the-art results across multiple benchmarks by framing all NLP tasks as text-to-text problems \cite{raffel2019exploring}. 

Even with these models' success, there are still problems. There is still a lot of research and discussion going on regarding topics like model interpretability, data privacy, and the possibility of producing biased or harmful content. The pursuit of more resilient and moral AI systems has led to continuous endeavors to enhance the accountability and transparency of data-driven natural language processing models \cite{mitchell2019model, bender2021dangers}.

Furthermore, there's a growing interest in combining data-driven and symbolic approaches to take advantage of both paradigms' advantages. The goal of hybrid systems is to bring together the data-driven models' learning capabilities with the rule-based accuracy of symbolic methods. Hybrid techniques, for instance, have been applied to enhance neural networks' interpretability and reasoning powers, allowing them to function better on tasks requiring structured knowledge and logical inference \cite{tenney2019bert, yang2020joint}.

Novel approaches for transforming natural language into data visualizations have been investigated in the context of recent developments in the NL2VIS field. To produce visualizations from natural language queries, Maddigan and Susnjak's Chat2VIS system makes use of large language models (LLMs) such as ChatGPT, Codex, and GPT-3 \cite{chat2vis}. This method addresses the inherent ambiguities and underspecifications of natural language by uniquely using prompt engineering, providing a dependable end-to-end solution for creating visualizations. Traditional NL2VIS systems, which frequently rely on specially designed models and manually crafted grammar rules, are much more expensive and complex than Chat2VIS. 

Chat2VIS can interpret user queries and produce visualizations with high accuracy and efficiency thanks to the use of pre-trained LLMs. The system does have certain drawbacks, though. The non-deterministic nature of LLMs leads to variability in plot generation, and there is room for improvement in the consistency of plot aesthetics, such as background colors and grid lines.

\begin{figure*}[!t]
    \centering
    \captionsetup{justification=centering}
    \includegraphics[width=0.98\textwidth]{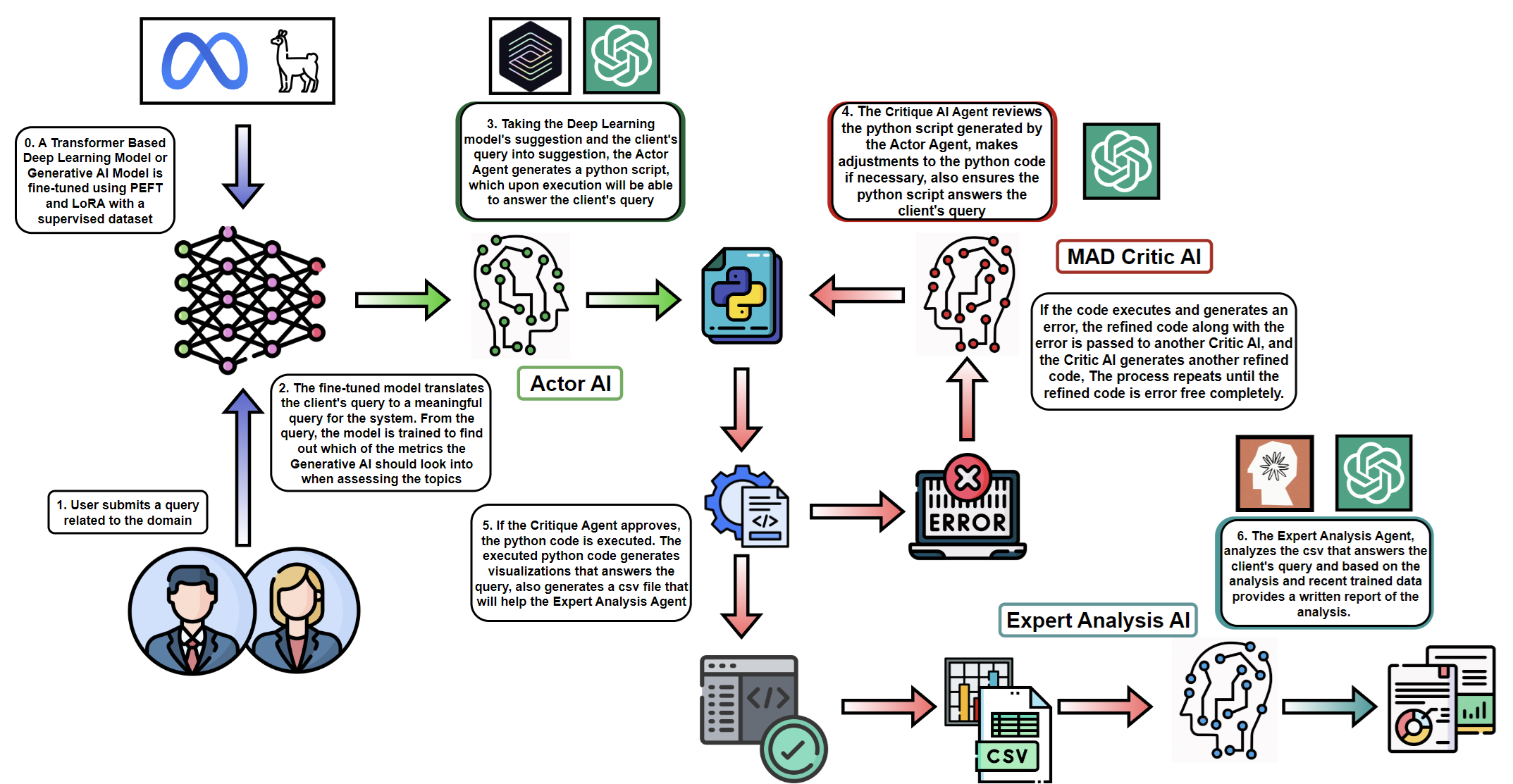}
    \caption{Framework for Multi-Agent Strategic Query Resolution and Analytical Diagnostic Tool (MASQRAD)}
    \label{fig:framework}
\end{figure*}

\subsection{Multi-Agent AI Systems Overview} In multi-agent AI systems, several intelligent agents collaborate to accomplish challenging tasks. Each agent acts independently based on both local and collective information \cite{wooldridge2009introduction}. These systems are essential in domains that call for decentralized decision-making and problem-solving. For example, distributed data is used by federated learning systems to facilitate collaborative learning without sacrificing privacy \mbox{\cite{konevcny2016federated}}. Agents in reinforcement learning multi-agent systems (RL-MAS) efficiently adapt to changing environments by learning the best course of action through trial and error \mbox{\cite{busoniu2008comprehensive}}. By mimicking natural processes, swarm intelligence models—like particle swarm optimization and ant colony optimization—solve optimization problems and exhibit strong problem-solving abilities by cooperatively exploring a variety of solutions \mbox{\cite{kennedy1995particle, dorigo2019ant}}. For automated negotiation and resource allocation, negotiation-based models handle conflict resolution and agreement-making between agents with different goals \mbox{\cite{kraus1997negotiation, fatima2002multi}}. By combining policy-based and value-based reinforcement learning techniques, actor-critic models stand out among these for complex query resolution tasks. In these models, the critic assesses the actions proposed by the actor, allowing for effective and flexible decision-making \mbox{\cite{konda2000actor}}. Because of its capacity to iteratively improve and optimize responses based on ongoing evaluative feedback, this architecture works especially well in complex query resolution tasks, offering a competitive edge in settings that require a high degree of accuracy and flexibility.

Our suggested framework incorporates a multi-agent actor-critic model that blends sophisticated deep learning techniques with symbolic reasoning, building on Chat2VIS's strengths. By addressing the shortcomings of both data-driven and symbolic approaches, this hybrid strategy seeks to offer a more reliable answer for data analysis and query resolution. In contrast to Chat2VIS, our system uses a Critic Generative AI to examine and improve the generated scripts, guaranteeing their accuracy and efficiency. The contextual analysis and thorough reports offered by the Expert Analysis Generative AI further enhance the overall practicality and usefulness of the visualizations. An analysis of the systems discussed is presented in a qualitative comparison in Table \ref{tab:comparison}.

\section{Methodology}

The Multi-Agent Strategic Query Resolution and Analytical Diagnostic (MASQRAD) System was developed and implemented using several different processes and techniques, all of which are detailed in the methodology section. To provide reliable query resolution, generate Python scripts for data visualization, validate and improve these scripts, and generate thorough analytical reports, the system uses sophisticated transformer-based models and a multi-agent framework. Four important subsections comprise the methodology:

\vspace{-0.1in}
\subsection{Query Interpretation with RoBERTa and LLaMA}

\begin{figure*}[!t]
    \centering
    \captionsetup{justification=centering}
    \includegraphics[width=0.98\textwidth]{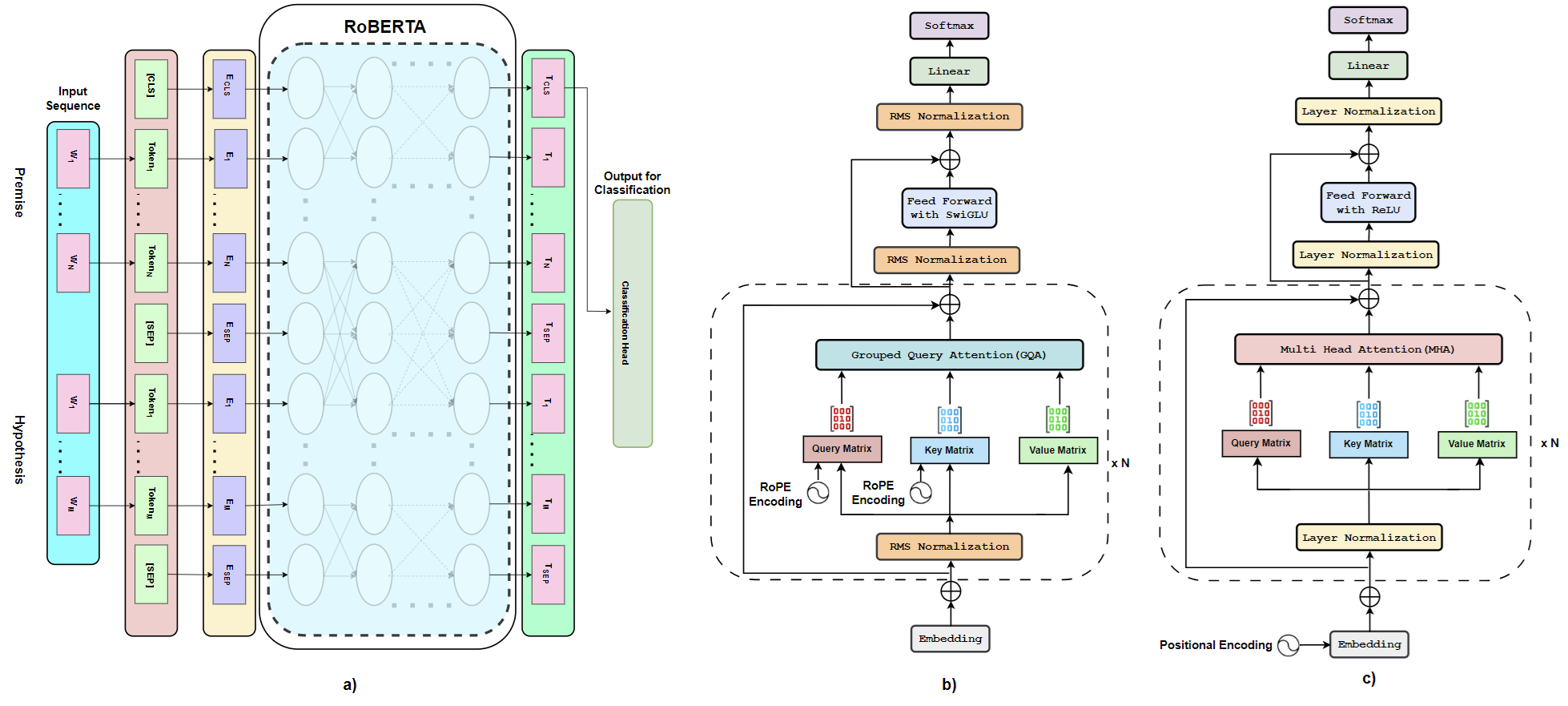}
    \caption{Basic Architecture Diagram of a) RoBERTa, b) Llama and c) GPT }
    \label{fig:llm_archi}
\end{figure*}

\subsubsection{Overview of RoBERTa}
RoBERTa ("Robustly Optimized BERT Approach") modifies the original BERT architecture by enhancing training efficiency and removing the next-sentence prediction task. This model, specifically the RoBERTa large, employs 24 layers, each with 1024 hidden units and 16 self-attention heads, totaling about 355 million parameters. \cite{liu2019roberta}. The basic architecture diagram of RoBERTa is illustrated in Figure \ref{fig:llm_archi}(a). In our system, RoBERTa excels at transforming ambiguous user queries into structured, actionable clues through a multilabel classification framework, enabling precise guidance for subsequent AI operations.

\subsubsection{Application of RoBERTa in Our System}
As a key component of our system, RoBERTa translates intricate queries into targeted clues that direct the Actor AI's response tactics. RoBERTa is used to predict multiple relevant clues from each query by utilizing its refined capabilities. These clues or indicators correlate to various aspects of the query, giving the Actor AI a complete set of guidelines for proceeding efficiently.

The model processes input queries and predicts the likelihood of each metric being relevant, which is formulated as follows:
\begin{equation}
\textbf{P}_{\text{labels}} = \sigma(\textbf{W} \cdot \text{RoBERTa}(\textbf{x}) + \textbf{b})
\end{equation}
where \( \textbf{x} \) is the tokenized input query, \( \textbf{W} \) is a weight matrix that maps RoBERTa's output embeddings to the label space, \( \textbf{b} \) is a bias vector, and \( \sigma \) is the sigmoid activation function applied element-wise. This function outputs probabilities between 0 and 1, representing the likelihood of each metric's relevance to the query.

Using this method, RoBERTa can provide the Actor AI with accurate, probabilistically weighted insights that facilitate sophisticated decision-making and customized responses based on the unique requirements of the user.
\\
\subsubsection{Overview of LLaMA}
Equipped with 13 billion parameters, LLaMA-2-13b presents novel processing improvements such as the Grouped Query Attention (GQA) method. The goal of this model (Figure \ref{fig:llm_archi}(b)) is to optimize computational resources and contextual responsiveness by grouping and selectively attending to the most relevant portions of the input.

The GQA mechanism is formulated as follows:
\\

\begin{equation}
\text{GQA} = \text{softmax}\left(\frac{\textbf{Q}W^Q \circ (\textbf{K}W^K)^T}{\sqrt{d_k}}\right) \odot (\textbf{V}W^V)
\end{equation}

In this equation:
\begin{itemize}
\item \(\textbf{Q}, \textbf{K}, \textbf{V}\) represent the query, key, and value matrices, respectively, which are standard components in attention mechanisms.
\item \(W^Q, W^K, W^V\) are the parameter matrices that transform the input query, key, and value data into higher-dimensional spaces for better feature extraction.
\item The element-wise multiplication \(\circ\) between \(\textbf{Q}W^Q\) and \((\textbf{K}W^K)^T\) allows the model to focus on how closely each element of the query aligns with elements of the key, enhancing the specificity of the attention mechanism.
\item \(\sqrt{d_k}\) is a scaling factor used to avoid overly large values during matrix multiplication that can lead to gradient vanishing problems.
\item The softmax function is applied to normalize the attention scores, ensuring they sum to one, thus representing probabilities.
\item The final element-wise multiplication \(\odot\) between the normalized attention scores and \(\textbf{V}W^V\) weights the value matrix, emphasizing the parts of the data that are most relevant to answering the query.
\end{itemize}

By concentrating computational efforts on the most notable elements of the input data, this sophisticated attention mechanism enables LLaMA to process multiple queries at once, improving efficiency in handling complex information scenarios. Additionally, Llama encodes the positional information of tokens using Rotary Positional Embeddings (RoPE).

\begin{equation}
\text{RoPE}_{pos} = \textbf{pos} \cdot \cos(\omega_{pos}) + \textbf{pos} \cdot \sin(\omega_{pos})
\end{equation}

where $\omega_{pos}$ represents the frequency of the positional encodings, enabling the model to preserve the relational context across different parts of the sequence.

\subsubsection{Application of LLaMA in Our System}
By providing imaginative, contextually rich interpretations of the same questions, LLaMA complements RoBERTa. It functions autonomously, generating creative recommendations that broaden and deepen the Actor AI's analytical capabilities and improve the system's capacity to create thorough and varied responses.

\subsection{Actor Agent-Driven Python Script Generation}

The GPT (Figure \ref{fig:llm_archi}(c)) family of advanced Generative AI models, namely GPT-3.5 Turbo and Codex, are employed by the Actor Agent in our system. These models,  based on OpenAI's transformer architecture, are essential for producing Python scripts that dynamically react to user queries.

\subsubsection{GPT-3.5 Turbo and Codex as Actor AI}
With a sophisticated multi-head self-attention mechanism that helps them manage complex data dependencies, GPT-3.5 Turbo and Codex both build on the capabilities of GPT-3. Because our system requires dynamic responses, this architecture makes it easier to generate contextually rich text based on input prompts \cite{brown2020language, vaswani2017attention}.

\begin{itemize}
    \item \textbf{Multi-Head Attention (MHA):} Central to their functionality, MHA allows these models to assess and synthesize different aspects of the input data through multiple "heads" of attention.
    \begin{equation}
    \text{MHA}(\textbf{Q}, \textbf{K}, \textbf{V}) = \text{Concat}(\text{head}_1, \dots, \text{head}_h)W^O
    \end{equation}
    \begin{equation}
    \text{head}_i = \text{softmax}\left(\frac{(\textbf{Q}W_i^Q)(\textbf{K}W_i^K)^T}{\sqrt{d_k}}\right)\textbf{V}W_i^V
    \end{equation}
    Where \(W_i^Q, W_i^K, W_i^V\) are the parameter matrices for the query, key, and value in each head, and \(W^O\) is the output transformation matrix.
    \item \textbf{Positional Encoding:} Positional encodings are integrated into input embeddings to maintain the sequence order, enhancing the model's ability to process language \cite{vaswani2017attention}.
\end{itemize}

\subsubsection{Mitigating Hallucination Error}
Hallucination errors, where models generate non-factual or irrelevant content, are a significant concern with language generation models. To mitigate such errors, the Actor Agent leverages contextual cues and structured inputs from RoBERTa and LLaMA, ensuring that the generated Python scripts remain accurate and relevant to the provided data \cite{xu2023understanding}.

\subsubsection{Functionality of the Actor Agent}
The primary role of the Actor Agent is to generate Python scripts that:
\begin{itemize}
    \item Produce data visualizations to answer user queries through intuitive graphs and charts.
    \item Perform numerical analysis and store results in dataframes for subsequent examination by the Expert Analysis Agent.
\end{itemize}
These scripts are tailored based on insights from RoBERTa and LLaMA, guided by the robust AI capabilities of GPT-based models, ensuring precise and actionable outputs.

The incorporation of these models improves the system's capacity to provide intelligent and tailored answers efficiently, enabling an advanced analytical procedure to handle intricate inquiries.

\subsection{MAD Critic Agent’s Script Validation and Refinement}

The Critic AI, a pivotal component of our system, employs the GPT-4-turbo model to review and refine the Python scripts generated by the Actor AI. This step is crucial to ensure the scripts not only execute without errors but also accurately address the user's queries through appropriate data visualizations and analytics.

\subsubsection{Functionality of the Critic AI}
The primary role of the Critic AI is to meticulously review the Python scripts produced by the Actor AI. This involves several checks and enhancements:
\begin{itemize}
    \item \textbf{Validation of Script Execution:} The Critic AI first verifies that the script can execute successfully and is capable of generating the required graphs and dataframes relevant to the original user query. This step ensures that the output precisely aligns with what is necessary to answer the user's query effectively.
    \item \textbf{Storage and Accessibility:} It ensures that all outputs, including charts and dataframes, are stored in appropriate cloud locations, accessible to the Expert Analysis Agent for further processing and analysis.
    \item \textbf{Enhancement of Outputs:} Beyond functional validation, the Critic AI enhances the aesthetic aspects of the graphs and optimizes the performance of the script, improving execution efficiency and visual appeal.
\end{itemize}

\subsubsection{Critic AI Powered by Multi Agent Debate}
Once validation has been completed successfully, the script is run to produce the required results. Execution errors, on the other hand, send the script to a different instance of Critic AI along with the error message, dataset URL, and user query. In response, several instances of Critic AIs begin an iterative process of refining the script in a Multi Agent Debate (MAD):

\begin{itemize}
    \item \textbf{Iterative Refinement:} Each Critic AI attempts to correct any errors identified by its predecessors, rewriting the script if necessary to ensure error-free execution.
    \item \textbf{Multi Agent Debate:} This process embodies a robust debate among several AI agents, each critiquing and enhancing the script to not only resolve errors but also to optimize the script’s overall effectiveness and accuracy.
\end{itemize}

In order to improve accuracy and reliability in automated systems, this iterative debate process leverages collective AI capabilities to gradually refine responses, which is advantageous for complex query resolution. 

The Critic AI's capacity to produce flawless and highly refined outputs is greatly improved by the integration of Multi Agent Debate into its workflow, guaranteeing that the end-user gets the most accurate and perceptive data possible \cite{smit2023we}.

\subsection{Expert Analysis and Report Generation}

As a key component of our artificial intelligence system, the Expert Analysis Agent generates comprehensive analysis reports and analyzes dataframes after the python script is executed. Claude-3.5 Sonnet \cite{bai2022constitutional,claude2023} and GPT-4-omni \cite{openai2024gpt4o} models have been tested in our system. Even though both models have demonstrated efficacy, the Claude-3.5 Sonnet is usually chosen because of its superior analytical capabilities and graduate-level reasoning, which enable it to handle highly complex analytical tasks with exceptional skill \cite{anthropic2023claude35}.

\subsubsection{Functionality of the Expert Analysis Agent}
The Expert Analysis Agent's primary function is to process the numerical results stored in dataframes, which represent the outcomes of Python scripts executed by the Actor Agent. The responsibilities of the Expert Analysis Agent include:
\begin{itemize}
    \item Analyzing dataframes to extract meaningful insights and patterns that directly respond to the user's queries.
    \item Generating comprehensive reports that not only detail the findings from the data but also provide contextually relevant insights derived from the latest information available to GPT-4-omni and Claude 3.5 Sonnet.
    \item Enhancing the understanding of the visual representations, such as graphs and charts, by correlating them with numerical analyses to provide a thorough interpretation of the results.
\end{itemize}

\subsubsection{Insight Generation Powered by Advanced AI}
Utilizing the advanced capabilities of Claude-3.5 Sonnet and GPT-4-omni, the Expert Analysis Agent synthesizes the insights from the data with contemporary knowledge extracted from the internet. This integration allows the Agent to:
\begin{itemize}
    \item Deliver insights that are not only based on the raw data but are also enhanced by the most recent and relevant information, making the analyses more valuable and actionable for the user.
    \item Provide recommendations and strategic advice that are informed by the latest trends and data, ensuring that the analysis is not only accurate but also aligned with current real-world applications.
\end{itemize}

The use of sophisticated AI models ensures that the Expert Analysis Agent is equipped to handle complex analytical tasks and generate reports that are both informative and practically useful for decision-making processes.

\section{Experiment Settings and Results}

The experimental setup that was utilized to set up and train our system's interpreter models, RoBERTa and Llama, is described in this section. Understanding how these models are ready to handle and interpret complex queries requires an understanding of these settings. In order to maximize the performance of all generative AI models, this section will also introduce the Prompt Engineering strategies. Using our MASQRAD system, we have validated its performance at the end of this section by testing it on a subset of the NL4DV and nvBench datasets.

\subsection{Experiment Settings for Interpreter Models}

\subsubsection{Experiment Settings for RoBERTa}

The main objective of training the RoBERTa models was to improve their Query Interpreter skills so that they could offer helpful suggestions for efficiently responding to user inquiries. To do this, we optimized pre-trained RoBERTa models by fine-tuning the model on a multilabel classification task, in which the model predicts particular clues or indicators on which to build its responses to questions pertaining to a given dataset. Because of this, every dataset required a different RoBERTa model to be trained, so every model became an expert in its own domain. The following training methods and configurations were used for RoBERTa: \\

\begin{figure*}[!t]
    \centering
    \captionsetup{justification=centering}
    \includegraphics[width=0.98\textwidth]{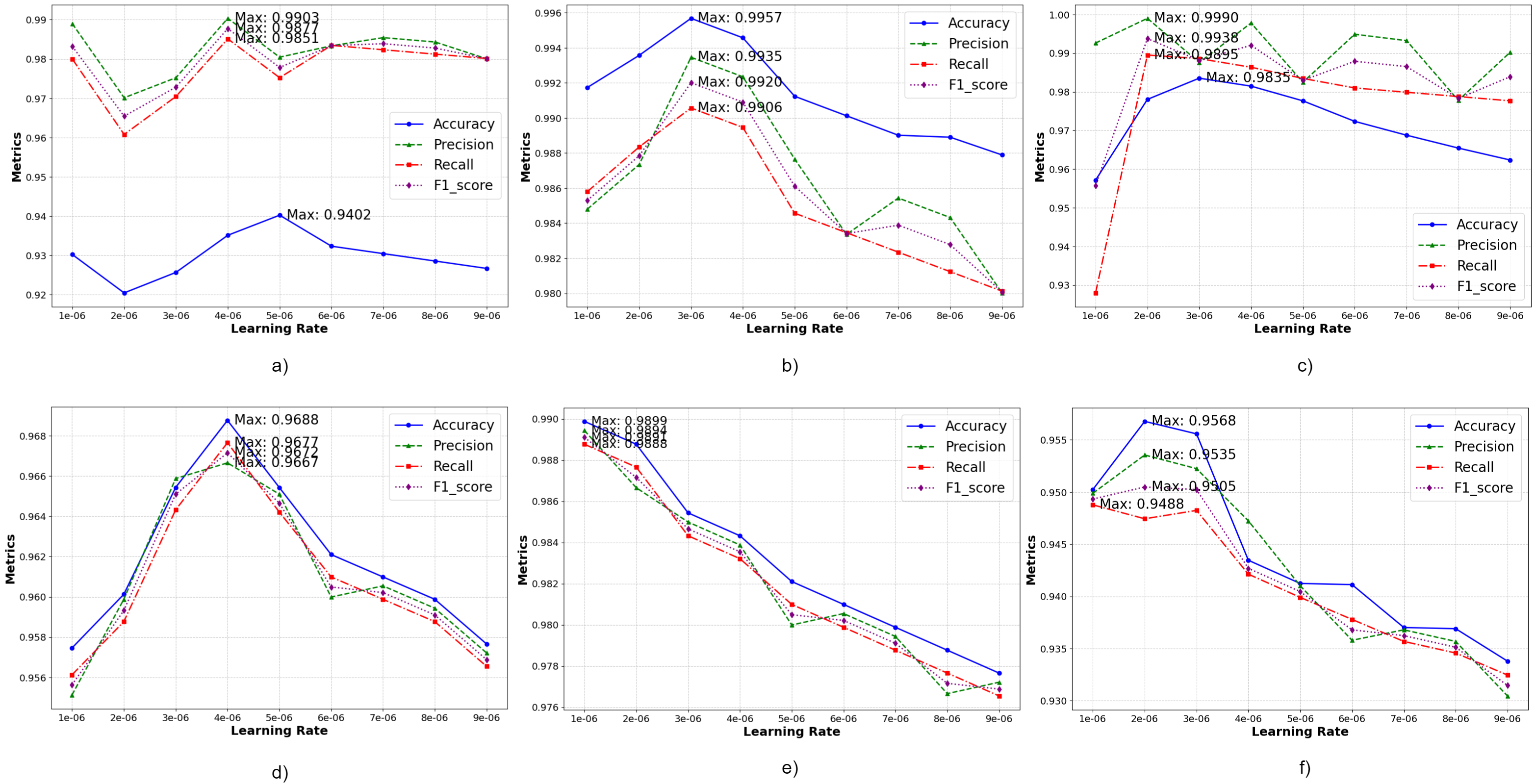}
    \caption{Performance Results of the trained interpreter RoBERTa models on the test query datasets for the (a) Movies Dataset (nvBench + NL4DV), (b) Cars Dataset (nvBench + NL4DV), (c) Superstore Dataset (NL4DV), (d) Inn Dataset (nvBench), (e) Euro Players Dataset (NL4DV) and (f) Architecture Dataset (nvBench)}
    \label{fig:framework}
\end{figure*}

\noindent \textbf{Dataset Preparation:} In order to effectively train our RoBERTa interpreter models to handle imprecise queries and translate them into clear, concise queries, we created several customized query datasets that were specific to each dataset we used for our analyses. This was imperative to make sure the models were ready to handle and interpret queries unique to various contexts. To test our MASQRAD system, for example, we created a special query dataset for the Movies category, which we then used to analyze the \textit{"movie\_1"} and \textit{"cinema"} databases from the nvBench dataset and the \textit{"Movies"} dataset from NL4DV. The utilization of Generative Language Models guided the creation of these query datasets, guaranteeing that any data present in the corresponding datasets might be applied to address the queries. 1000 queries were included in each query dataset to train the multilabel classification models. The quantity of ground truth labels differed based on the particular dataset under consideration. It is significant to highlight that the unbalanced nature of these datasets created extra difficulties for the training process. Table \ref{tab:query_datasets} summarizes these query datasets and sheds light on the variety and complexity of the data that our system was trained on. This dataset goes through a number of preprocessing stages:
\begin{table}[h]
\centering
\begin{tabular}{|l|l|l|l|}
\hline
\textbf{Query} & \textbf{Target} & \textbf{Source} & \textbf{No. of} \\
\textbf{Dataset} & \textbf{Dataset} & \textbf{Benchmark} & \textbf{Query} \\
\textbf{Categories}   & \textbf{Name}            & \textbf{Name}             & \textbf{Samples}      \\ \hline
\multirow{3}{*}{Movies} & movie\_1 & nvBench & \multirow{3}{*}{1000} \\
                        & cinema   & nvBench &  \\
                        & Movies   & NL4DV   &  \\ \hline
\multirow{2}{*}{Cars}   & car\_1   & nvBench & \multirow{2}{*}{1000} \\
                        & Cars     & NL4DV   &  \\ \hline
Superstore              & Superstore & NL4DV  & 1000 \\ \hline
Inn                     & inn\_1     & nvBench & 1000 \\ \hline
Euro Players            & Euro       & NL4DV   & 1000 \\ \hline
Architecture            & architecture & nvBench & 1000 \\ \hline
\end{tabular}
\caption{Summary of Query Datasets Used for Training and Testing the Interpreter Models}
\label{tab:query_datasets}
\end{table}

    \begin{itemize}
        \item Tokenization using RobertaTokenizer, configured for a maximum length of 64 tokens.
        \item Shuffling and resetting the index to ensure an unbiased order of input data.
    \end{itemize}

\noindent  \textbf{Model Configuration:}
    \begin{itemize}
        \item Model: Roberta Large Model from the transformers library.
        \item The model employs the pooled embedding from the first token tensor of the last hidden state for classification, effectively capturing the contextual essence of each query.
    \end{itemize}
    
\noindent    \textbf{Training Process:}
    \begin{itemize}
        \item The dataset was divided into training, validation, and testing segments in the ratios of 50\%, 20\%, and 30\% respectively, to monitor learning progress and prevent overfitting.
        \item The model was trained on a batch size of 16, and for 40 epochs, using the Adam optimizer with a learning rate ranging from \(1 \times 10^{-6}\) to \(9 \times 10^{-6}\), to optimize binary cross-entropy loss.
    \end{itemize}

These settings aim to fine-tune RoBERTa for efficient textual query understanding and processing, which is essential for the system's functionality in real-world situations. Because of this setup, RoBERTa can be easily incorporated into our larger system architecture, improving the AI platform's overall accuracy and responsiveness. \\

\noindent \textbf{Multilabel Classification Performance:} The experiments conducted on the interpreter RoBERTa models across various datasets demonstrated robust performance, with key evaluation metrics varying slightly depending on the dataset and learning rate applied. Across all models, the metrics observed ranged as follows:

\begin{itemize}
    \item \textbf{Accuracy:} Across all datasets, the accuracy metric maintained a high level of consistency, with values ranging from roughly 93\% to slightly less than 100\%. 

    \item \textbf{Precision and Recall:} While recall and precision metrics performed well, they were generally marginally worse than accuracy. These metrics, which showed the models' efficient relevance and retrieval of query-related metrics, typically fell between 92\% and 99\%.
 
    \item \textbf{F1 Score:} The F1 scores, which take precision and recall into account, were usually between 94\% and 99\%. This highlights the models' capacity to stay balanced in their performance even after fine-tuning to particular query datasets.
\end{itemize}

Lower learning rates, from $2 \times 10^{-6}$ to $5 \times 10^{-6}$, generally fostered more stable convergence in metrics, with optimal performance often observed at learning rates ranging from $1 \times 10^{-6}$ to $9 \times 10^{-6}$. The responses of each dataset to changes in the learning rate indicate that optimizing model effectiveness requires a customized approach to parameterization for each domain. 

These findings validate the effectiveness of RoBERTa models in the MASQRAD system for deciphering and converting ambiguous queries into useful insights, especially when optimized on query datasets specific to a given domain.

\subsubsection{Experiment Settings for Llama}

Because of its text generation capabilities, the Llama model—in particular, the Llama-2-13b-hf version—has been used. This model is a component of the larger architecture of our system that helps it understand and react to complex queries. 

\begin{itemize}
    \item \textbf{Model Configuration:}
    \begin{itemize}
        \item \textbf{Model Setup:} LlamaForCausalLM, a model optimized for text generation, was loaded from a pre-trained configuration specifically designed for high-fidelity and contextually aware text outputs.
        \item \textbf{Tokenizer:} LlamaTokenizer was used to ensure that text inputs are appropriately converted into token formats that are compatible with the model, supporting a maximum new token generation of 64 tokens per query.
    \end{itemize}

    \item \textbf{Query Handling:}
    \begin{itemize}
        \item A pipeline for text generation was established using the Llama model and tokenizer, configured to handle various data types and computational optimizations such as setting the torch data type to float16 for efficiency.
        \item The device map was automatically set to optimize resource allocation during model operation.
    \end{itemize}

    \item \textbf{Operational Details:}
    \begin{itemize}
        \item \textbf{Query Processing:} The model takes input queries and processes them to generate responses based on predefined indicators or hints to answer the query. 
        \item \textbf{Creative Responses:} To ensure that the output is concentrated on offering the most important information, the model is prompted by each query to selectively generate text based on the most relevant clues. The output from this module is similar to what the RoBERTa models produce, in which domain-specific knowledge from previous training greatly influences the outputs. On the contrary,  the Llama model can provide unconventional recommendations or clues that deviate from the structured outputs of RoBERTa since its responses are instantaneous and not influenced by particular dataset knowledge. 
        \item \textbf{Clues Extraction:} A method to extract relevant clues from the generated responses was implemented, using a regex pattern to match and collect responses associated with specific performance indicators.
    \end{itemize}
\end{itemize}

For our system to effectively interpret user queries and provide contextually relevant and accurate responses, the Llama model setup is essential. Because of its exceptional ability to produce accurate and contextually aware responses, the Llama model is a crucial part of our query resolution system. 

\begin{figure*}[!t]
    \centering
    \captionsetup{justification=centering}
    \includegraphics[width=0.8\textwidth]{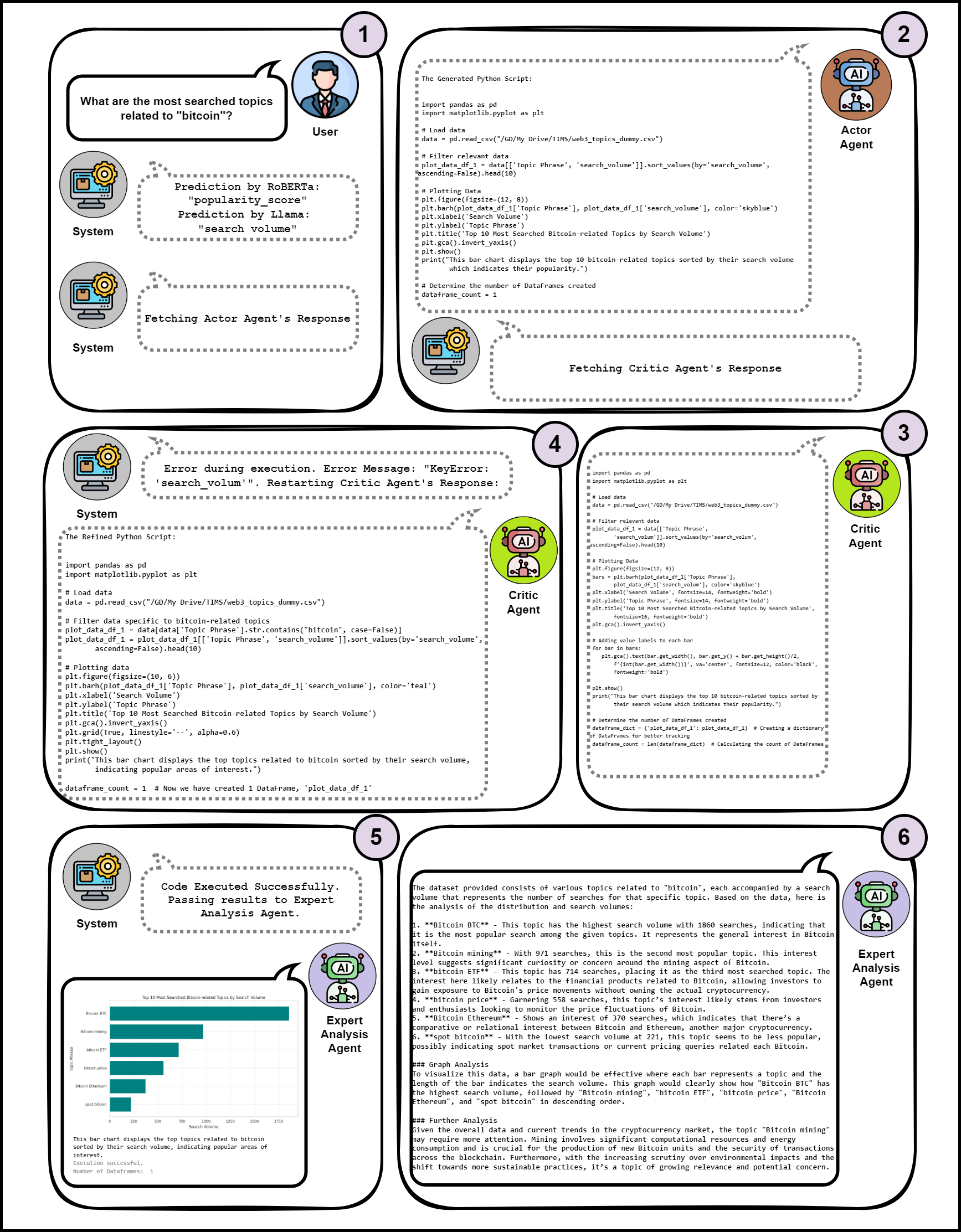}
    \caption{Example of a Query Resolution by the MASQRAD System}
    \label{fig:masqrad_demo}
\end{figure*}

\subsection{Prompt Engineering for Generative LLMs}

In order to ensure that the responses from generative language models are accurate and specifically tailored to the operational requirements at various stages of query resolution, prompt engineering is a fundamental technique in our system \cite{liu2021pretrain, gao2021making, lester2021power}. Every prompt is designed with care to correspond with the desired results, regardless of whether that means following strict dataset limitations or utilizing more comprehensive sources of data.

\subsubsection{Focused Prompts for Actor and Critic Stages}
In the Actor and Critic phases, the prompts are formulated to guarantee that the generative models generate outputs that strictly adhere to the dataset's context. All outputs are guaranteed to be directly applicable to the dataset at hand thanks to this targeted prompting strategy, which reduces the possibility of producing irrelevant or out-of-scope content.

\begin{itemize}
    \item \textbf{Dataset-Specific Prompts:} For these stages, prompts specifically instruct the models to generate Python scripts or validate outputs based solely on the dataset provided. A generalized example might be, "Using only the data provided in \texttt{dataset\_url}, generate a Python script to analyze \texttt{column\_names} and produce visualizations that answer the user's query."
    \item \textbf{Containment Strategies:} These prompts ensure that the model doesn't hallucinate or incorporate irrelevant data by restricting the AI's response scope to the dataset, protecting the accuracy and significance of the outputs that are produced.
    \item \textbf{Conflict Resolution:} In the rare instances where conflicts arise, especially during the Critic AI phase, a multi-agent debate mechanism ensures resolution through consecutive agreement between Critic AIs, effectively preventing any single error or discrepancy from compromising the system's outputs.

\end{itemize}

\subsubsection{Expansive Prompts for Llama and Expert Analysis Stages}
On the other hand, prompts in the Llama and GPT for Expert Analysis stages are designed to take advantage of the models' ability to incorporate broader contextual information. By incorporating insights that go beyond the immediate dataset, this approach improves the analyses' depth and usefulness.

\begin{itemize}
    \item \textbf{Incorporating External Knowledge:} Prompts in these stages encourage models to draw on external sources and their pre-trained knowledge. For example, an Expert Analysis prompt might be, "Based on the data in \texttt{dataset\_url} and your broader knowledge, provide insights and potential action points that could benefit the user's understanding of the analyzed trends."
    \item \textbf{Utilizing Model Capabilities:} This strategy utilizes the full capabilities of the AI to enrich the analysis with internet-sourced information, offering more comprehensive insights and recommendations based on up-to-date trends and data.

\end{itemize}

\subsubsection{Ensuring Alignment with System Goals}
In order to match the responses of AI models with the strategic objectives of the system, such as raising decision-making efficiency by drawing insights from the generated results, prompt engineering is essential at every step. The generative AI agents, produce outputs that are both practically and technically valuable thanks to strategic prompt design.

\begin{itemize}
    \item \textbf{Strategic Alignment:} Each prompt is not only a directive for generating specific types of outputs but also a tool for ensuring that these outputs serve the broader objectives of the system effectively.
    \item \textbf{Agent Interaction:} Agents within the system share computational resources and storage, operating in a sequential process where the output of one agent serves as the input for another. This tightly integrated workflow minimizes the potential for conflicts, as each agent is designed to build upon the previous agent's results, ensuring coherent and unified progress towards the final output.
\end{itemize}

This sophisticated use of prompt engineering across different stages of our AI-driven query resolution system ensures that generative models produce high-quality, relevant, and actionable outputs, adhering closely to user needs and operational demands. Figure \ref{fig:masqrad_demo} illustrates an example of query resolution using the MASQRAD system, showcasing the comprehensive process from script generation, through script validation powered by multi-agent debate, to code execution, detailed analysis of generated visualizations, and the offering of insightful conclusions.

\subsubsection{Parameter Settings}

In our system, the generative AI agents are fine-tuned with specific parameter settings to optimize their performance for different tasks:

\textbf{Interpreter Models:} For the LLaMA model, which focuses on query interpretation, we use a conservative \textit{temperature} setting of \mbox{\(0.3\)} to prioritize precision and a \mbox{\(top\_p\)} value of \mbox{\(0.7\)} to balance diversity with relevance. The maximum number of new tokens generated is limited to \mbox{\(64\)}, ensuring concise yet comprehensive interpretations.

\textbf{Actor Agent:} The Actor Agent, tasked with generating Python scripts, operates with a \mbox{\(temperature\)} of \mbox{\(0.5\)} and \mbox{\(top\_p\)} of \mbox{\(0.9\)}. These settings enable creative but controlled script generation, maintaining a balance between inventiveness and adherence to the data context.

\textbf{Critic Agent:} To validate and refine scripts, the Critic Agent uses a \mbox{\(temperature\)} of \mbox{\(0.7\)} and \mbox{\(top\_p\)} of \mbox{\(0.8\)}. This slightly higher temperature allows for more rigorous scrutiny and innovative problem-solving during script reviews.

\textbf{Expert Analysis Agent:} The Expert Analysis Agent analyzes dataframes and generates reports with a \mbox{\(temperature\)} of \mbox{\(0.4\)} and \mbox{\(top\_p\)} of \mbox{\(0.6\)}, settings that foster analytical depth while maintaining high accuracy in insights.

These tailored settings ensure that each agent effectively fulfills its role within the MASQRAD system, enhancing overall performance and reliability.

\subsection{Evaluation on nvBench and NL4DV Benchmarks}

\begin{figure*}[!t]
    \centering
    \captionsetup{justification=centering}
    \includegraphics[width=0.88\textwidth]{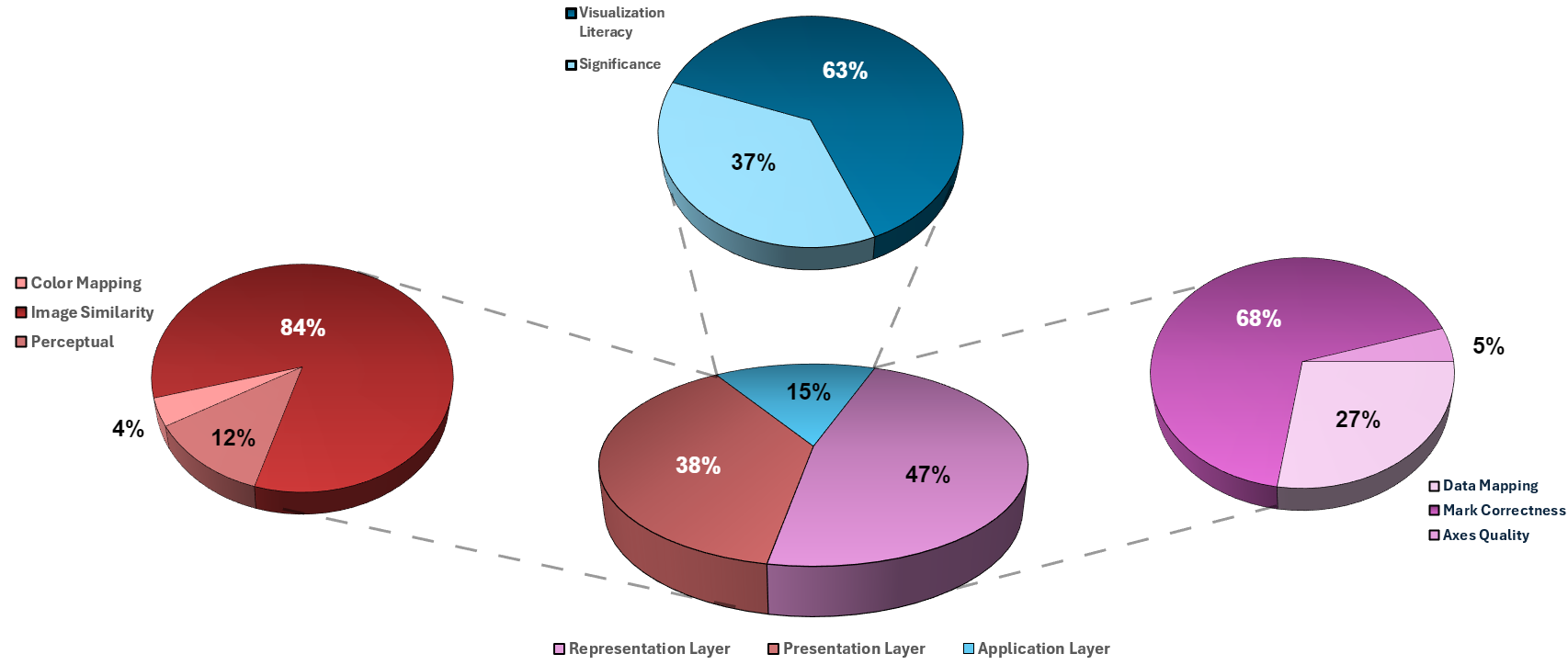}
    \caption{Types of inaccuracies identified in the MASQRAD system during evaluation with the Vi(E)va LLM stack on subsets of nvBench and NL4DV benchmarks, categorized by evaluation layers.}
    \label{fig:viva_llm_eval}
\end{figure*}

The selection of nvBench \mbox{\cite{luo2021nvbench}} and NL4DV \mbox{\cite{narechania2020nl4dv}} benchmarks was driven by their diverse and comprehensive datasets, which are widely recognized and utilized in the Natural Language to Data Visualization (NL2VIS) domain. These benchmarks include a variety of visualization challenges that reflect real-world scenarios, making them ideal for robustly testing MASQRAD's capabilities. Additionally, to efficiently evaluate our system without the need for excessive domain-specific model training, we carefully selected datasets that featured overlapping or similar domains across both benchmarks. This strategic choice allowed us to utilize a limited number of fine-tuned RoBERTa models, thereby optimizing our resource allocation and focusing on the most relevant and comparative analysis across benchmarks. Specifically, from the nvBench benchmark, we selected the datasets \textit{movie\_1, cinema, architecture, car\_1}, and \textit{inn\_1}. From the NL4DV benchmark, we chose the datasets \textit{Movies, Cars, Superstore,} and \textit{Euro}. The total number of queries processed from each of these datasets is detailed in Table \mbox{\ref{tab:performance_comparison}}, providing a clear overview of our comprehensive evaluation approach. 

The Vi(E)va LLM evaluation stack \mbox{\cite{podo2024vi}}, utilized to critically assess the visualizations produced by MASQRAD, consists of several layers that comprehensively evaluate different aspects of visualization quality and effectiveness. Although the stack includes additional layers such as the LLM Layer and Code Layer, our focus was on the Representation, Presentation, and Application Layers. These layers were specifically chosen as they directly assess the visual output, which is central to the functionality of MASQRAD. The LLM Layer, focusing on `LLM Effort,' and the Code Layer, evaluating `Grammar Similarity,' `Code Similarity,' and `Syntax Correctness,' were considered less critical for our initial evaluations as our primary goal was to measure the visual interpretation and applicability of the system outputs. The Representation Layer, Presentation Layer, and Application Layer of the Vi(E)va LLM evaluation stack were used to evaluate the produced visualizations critically. The functionalities and criteria of these layers are detailed below:

\subsubsection{Representation Layer}
This layer focuses on the accuracy and correctness of the data represented in the visualizations:
\begin{itemize}
    \item \textbf{Data Mapping:} Evaluates the data correctness of the generated visualizations compared to the ground truth.
    \item \textbf{Mark Correctness:} Assesses whether the visual marks between the ground truth and generated visualizations are accurate and consistent.
    \item \textbf{Axes Quality:} Measures the quality of the visualization axes, ensuring they align with the ground truth and the context of the user query.
\end{itemize}

\subsubsection{Presentation Layer}
This layer examines the aesthetic and perceptual quality of the visualizations:
\begin{itemize}
    \item \textbf{Color Mapping:} Assesses the quality of color mapping in relation to the type of data presented and the user query.
    \item \textbf{Image Similarity:} Compares the similarity of images between the generated and ground truth visualizations.
    \item \textbf{Perceptual Similarity:} Evaluates human perceptual accuracy in recognizing and understanding the visualizations.
\end{itemize}

\subsubsection{Application Layer}
This final layer evaluates the applicability and insightfulness of the visualizations:
\begin{itemize}
    \item \textbf{Visualization Literacy:} Assesses the visualizations against the best standards and practices in the field.
    \item \textbf{Significance:} Measures the insightfulness and significance of the information conveyed by the visualizations.
\end{itemize}

The MASQRAD system was evaluated by a team of online brand marketing experts using this structured evaluation framework. This ensures a comprehensive examination of all aspects of the visualizations, providing a thorough evaluation of their effectiveness and relevance in real-world situations. Discrepancies in one or more of the aforementioned evaluation criteria were found for the visualizations that were not precisely aligned with the ground truth. 

Table \ref{tab:dataset_performance} displays the total number of queries processed and the errors found in the results, providing an overview of the MASQRAD system's performance across multiple datasets. It is noteworthy that the level of insightful analysis produced by the Expert Analysis Agent of the MASQRAD system was not assessed in this study. Since there are currently no NL4VIS systems with comparable capabilities or established benchmarks for these assessments, no comparative analysis were made regarding this part of the system. 

\begin{table}[htb]
\centering
\begin{tabular}{|l|l|c|c|}
\hline
\textbf{Dataset}    & \textbf{Benchmark} & \textbf{\makecell{Total \\Queries}} & \textbf{\makecell{Inaccurate \\ Responses}} \\ \hline
Movies              & NL4DV               & 39                     & 4                     \\ \hline
movie\_1 + cinema   & nvBench             & 103                    & 22                    \\ \hline
architecture        & nvBench             & 22                     & 6                     \\ \hline
Cars                & NL4DV               & 44                     & 5                     \\ \hline
car\_1              & nvBench             & 100                    & 11                    \\ \hline
Superstore          & NL4DV               & 23                     & 3                     \\ \hline
inn\_1              & nvBench             & 156                    & 12                    \\ \hline
Euro                & NL4DV               & 13                     & 1                     \\ \hline
\textbf{Total}      & -                   & \textbf{500}           & \textbf{64}           \\ \hline
\end{tabular}
\caption{Performance of the MASQRAD system on various datasets.}
\label{tab:dataset_performance}
\end{table}

\begin{table}[H]
\centering
\begin{tabular}{|l|c|}
\hline
\textbf{Approaches} & \textbf{Accuracy (\%)} \\
\hline
Seq2Vis \cite{luo2021synthesizing} & 2 \\
Transformer \cite{vaswani2017attention} & 3 \\
ncNet \cite{luo2021natural} & 26 \\
Chat2Vis \cite{chat2vis} & 43 \\
RGVisNet \cite{song2022rgvisnet} & 45 \\
Li et al. (Zero Shot) \cite{li2024visualization} & 43 \\
Li et al. (Few Shot) \cite{li2024visualization} & 50 \\
\textbf{MASQRAD} & \textbf{87} \\
\hline
\end{tabular}
\caption{Comparative performance of MASQRAD against other visualization approaches.}
\label{tab:performance_comparison}
\end{table}

A subset of the nvBench and NL4DV benchmarks were used to assess our MASQRAD system, with a sample size of 500 queries. Of these, 64 queries had errors in various Vi(E)va LLM stack evaluation criteria, as shown in Figure \ref{fig:viva_llm_eval}; this led to an accuracy rate of 87\% overall. The advanced capabilities of the GPT-4 Turbo models and the methodical, iterative refinement process used to create and optimize the code for visualizations are largely responsible for this high degree of accuracy. 

\begin{table}[htb]
\centering
\renewcommand{\arraystretch}{1.2}
\setlength{\tabcolsep}{2pt}
\caption{Inaccuracy Distribution across Different Layers and Subcategories for Each Dataset}
\label{tab:inaccuracy_distribution}
{\fontsize{6.5pt}{7.5pt}\selectfont 
\begin{tabular}{|l|c|c|c|c|c|c|c|c|c|}
\hline
\multirow{2}{*}{\textbf{Dataset}} & \multicolumn{3}{c|}{\textbf{Representation}} & \multicolumn{3}{c|}{\textbf{Presentation}} & \multicolumn{2}{c|}{\textbf{Application}} & \multirow{2}{*}{\textbf{Total}} \\ \cline{2-9}
                                  & \makecell{\textbf{Data}\\\textbf{Map.}} & \makecell{\textbf{Mark}\\\textbf{Corr.}} & \makecell{\textbf{Axes}\\\textbf{Qual.}} & \makecell{\textbf{Color}\\\textbf{Map.}} & \makecell{\textbf{Img.}\\\textbf{Sim.}} & \makecell{\textbf{Per-}\\\textbf{cept.}} & \makecell{\textbf{Vis.}\\\textbf{Lit.}} & \makecell{\textbf{Sig-}\\\textbf{nif.}} & \\ \hline
Movies                            & 1                  & 4                   & 0                   & 0                  & 4                 & 2                & 2                 & 0                & 13 \\ \hline
\makecell{movie\_1\\ + cinema}    & 2                  & 6                   & 9                   & 0                  & 8                 & 0                & 0                 & 0                & 25 \\ \hline
architecture                      & 0                  & 2                   & 0                   & 1                  & 0                 & 1                & 1                 & 1                & 6 \\ \hline
Cars                              & 1                  & 2                   & 0                   & 0                  & 1                 & 0                & 1                 & 0                & 5 \\ \hline
car\_1                            & 2                  & 3                   & 0                   & 1                  & 1                 & 1                & 2                 & 1                & 11 \\ \hline
Superstore                        & 1                  & 1                   & 0                   & 1                  & 0                 & 0               & 0                 & 0                & 3 \\ \hline
inn\_1                            & 2                  & 3                   & 0                   & 1                  & 1                 & 1                & 2                 & 2                & 12 \\ \hline
Euro                              & 1                  & 0                   & 0                   & 0                  & 0                 & 0                & 0                 & 0                & 1 \\ \hline
\textbf{Total}                    & \textbf{10}         & \textbf{21}         & \textbf{9}          & \textbf{4}         & \textbf{15}       & \textbf{5}       & \textbf{8}        & \textbf{4}       & \textbf{76} \\ \hline
\end{tabular}
}
\end{table}

Table \mbox{\ref{tab:inaccuracy_distribution}} presents a comprehensive breakdown of inaccuracies across different evaluation subcategories and datasets to give a more detailed understanding of the system's performance. This table provides information about possible areas for system improvement by highlighting the precise regions within each layer where inaccuracies were the most prevalent. We can find recurring trends and anomalies by classifying these errors, which is crucial for refining the MASQRAD system. It should be noted that the total number of inaccuracies (76) differs from the total number of queries with inaccuracies (64), as some queries exhibited errors across multiple evaluation categories.

For upcoming MASQRAD iterations, a number of mitigation techniques are suggested in response to the found errors. Particular attention will be given to improving natural language processing skills in order to handle ambiguous queries more effectively and increase the accuracy of data mapping. Furthermore, in order to guarantee greater consistency and accuracy in visualization outputs, developments in AI model training will concentrate on improving the inter-agent communication protocols and enriching the training datasets. It is anticipated that these tactics will greatly enhance the system's performance and dependability while addressing the underlying causes of the errors, especially those pertaining to mark correctness and data representation.

Several of the problems that Li et al.  \cite{li2024visualization} identified in their analysis with the nvBench benchmark also apply to our assessment. The ambiguity in the specification of visualization types in many queries was a significant issue that was observed; for example, in some queries, only pie charts were marked as correct in the ground truth when either pie chart or bar chart could be applicable. Other issues include inconsistent date formats in visualizations and incomplete queries, which have caused discrepancies between the generated and expected visualizations. Similar problems were found in the NL4DV dataset, which added to the errors noted. 

Moreover, although MASQRAD has been tested against a number of systems that have also been evaluated using nvBench or NL4DV, it is not apparent whether those systems were evaluated using all the datasets in these benchmarks. Table \ref{tab:performance_comparison} compares the performance of MASQRAD with these systems and demonstrates a significant advantage in achieving higher visualization accuracy.

\subsection{Scalability and Execution Times}

The MASQRAD framework showcases its scalability and efficient handling of diverse dataset structures and sizes, demonstrating consistent performance while adhering to API query limits.

\subsubsection{Scalability Features}
MASQRAD's architecture is designed around the schema of each dataset, which specifies the number of tables, rows, columns, and their types. This schema-driven approach enables the system to directly interact with the dataset's location, eliminating the need to transfer large volumes of data and ensuring efficient query processing without exceeding API query limits. The uniform response strategy, where the system consistently processes queries using a standardized approach regardless of dataset size, further enhances scalability and performance predictability.

\subsubsection{Performance Variability}
Execution times within the MASQRAD modules show variability, particularly in the Critic and Expert Analysis modules; however, this variability is not attributed to dataset sizes, as evidenced by the consistency across datasets in Table \mbox{\ref{tab:execution_times}}.

\begin{itemize}

 \item \textbf{Critic Response Times:} High standard deviations in execution times are primarily observed in the Critic module. This variability arises from the occasional activation of a Multi-agent debate, where agents engage in iterative refinements to optimize outputs, thereby extending the response duration.

 \item \textbf{Expert Analysis Complexity:} The Expert Analysis module sometimes experiences extended processing times, particularly when analyzing multiple visualizations generated for certain queries.

\end{itemize}

Table \mbox{\ref{tab:execution_times}} illustrates the execution times for MASQRAD's different modules across various datasets, emphasizing the framework's robustness and adaptability. This table also highlights instances where multi-agent interactions and complex analytical tasks prolong processing times, underscoring the system's capability to manage extensive and variable computational loads efficiently. The collaborative validation and optimization procedures in MASQRAD's multi-agent framework make it inherently more resource-intensive than single-agent systems. However, the notable increase in output accuracy and robustness justifies the extra computational expenses. In order to correct errors, single-agent systems frequently need extra post-processing or human intervention, which MASQRAD handles autonomously.


\begin{table}[htb]
    \centering
    \caption{Execution Times (in seconds) for MASQRAD Modules Across Datasets}
    \label{tab:execution_times}
    \renewcommand{\arraystretch}{1.2} 
    {\fontsize{6.5pt}{6.8pt}\selectfont  
    
    \begin{tabular}{|c|c|c|c|c|}
        \hline
        \multirow{2}{*}{\textbf{Dataset}} & \multicolumn{4}{c|}{\textbf{Module Execution Time (s)}} \\
        \cline{2-5}
        & \textbf{Module 1} & \textbf{Module 2} & \textbf{Module 3} & \textbf{Module 4} \\
        & \textbf{(Interpreters)} & \textbf{(Actor} & \textbf{(Critic with} & \textbf{(Expert} \\
        & & \textbf{Agent)} & \textbf{Multi-Agent} & \textbf{Analysis} \\
        & & & \textbf{Debate)} & \textbf{Agent)} \\
        \hline
        \makecell{movie\_1 + \\ cinema} & 31.45 $\pm$ 8.62 & 8.92 $\pm$ 4.81 & 15.45 $\pm$ 18.67 & 27.32 $\pm$ 11.65 \\
        \hline
        Movies & 33.21 $\pm$ 9.75 & 11.35 $\pm$ 5.44 & 16.02 $\pm$ 19.11 & 25.83 $\pm$ 13.07 \\
        \hline
        car\_1 & 30.88 $\pm$ 8.31 & 7.92 $\pm$ 4.25 & 17.81 $\pm$ 22.32 & 24.95 $\pm$ 11.98 \\
        \hline
        Cars & 28.74 $\pm$ 10.02 & 12.13 $\pm$ 5.92 & 19.43 $\pm$ 21.72 & 28.62 $\pm$ 14.78 \\
        \hline
        Superstore & 32.02 $\pm$ 9.14 & 9.83 $\pm$ 4.67 & 18.14 $\pm$ 22.95 & 27.77 $\pm$ 13.98 \\
        \hline
        inn\_1 & 29.48 $\pm$ 9.57 & 10.01 $\pm$ 5.11 & 16.53 $\pm$ 20.47 & 25.34 $\pm$ 12.28 \\
        \hline
        Euro & 31.93 $\pm$ 10.13 & 11.56 $\pm$ 6.07 & 17.32 $\pm$ 24.88 & 29.81 $\pm$ 15.12 \\
        \hline
        architecture & 28.67 $\pm$ 8.73 & 9.42 $\pm$ 4.92 & 15.92 $\pm$ 18.51 & 26.45 $\pm$ 12.64 \\
        \hline
    \end{tabular}
    }
    
\end{table}

\subsection{Implementation in Domain-Agnostic Setting}

\begin{figure}[htb]
    \centering
    \captionsetup{justification=centering}
    \includegraphics[width=0.70\columnwidth]{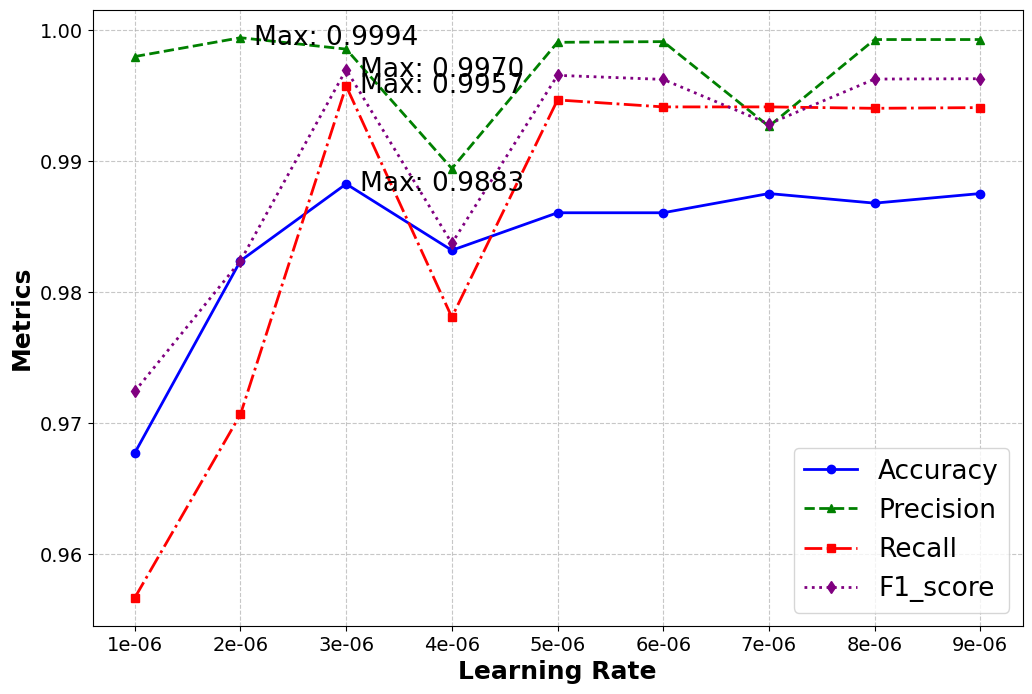}
    \caption{Performance of RoBERTa models trained on SEO Task across different learning rates}
    \label{fig:roberta_seo}
\end{figure}

The team of experts in online brand marketing also collaborated on a Search Engine Optimization (SEO) project where they integrated MASQRAD framework in their system. The project involved analyzing query datasets spanning diverse domains such as \textit{Web3, Metaverse, Vitality \& Health, Restoration,} and \textit{Capital Investment}. With 40 queries per domain, the MASQRAD framework was used to process 200 queries in total. The system's domain-agnostic capabilities were demonstrated by using the same fine-tuned RoBERTa model, which was trained on a multi-label classification task with 6800 queries and 52 labels, as the query interpreter across all domains. The performance of the trained interpreter model in domain-agnostic settings is shown in Figure \mbox{\ref{fig:roberta_seo}}. 

The marketing specialists used the Representation, Presentation, and Application layers of the Vi(E)va LLM evaluation stack to assess the visualizations and analyses generated by MASQRAD. To ensure consistency in assessment, the evaluation employed the same methodology as the benchmark datasets. A 69.5\% accuracy rate (139 correct visualizations generated out of 200 queries) showed that MASQRAD could handle a wide range of real-world query domains without domain-specific fine-tuning. 

This implementation demonstrates MASQRAD's versatility, showcasing its ability to deliver reliable insights and visualizations in domain-agnostic contexts. The collaborative nature of this project also shows how MASQRAD can be applied for data analysis and visualization problems in real-world industrial settings.

\section{Conclusion}

This paper introduced MASQRAD, a Multi-Agent Strategic Query Resolution And Diagnostic tool that enhances query resolution and data analysis through the synergistic interplay of multiple generative AI agents. MASQRAD translates vague user inputs into precise queries, generates tailored Python scripts, and provides detailed insights explaining the visualization results, distinguishing it from most NL2VIS systems which only generate visualizations.

MASQRAD demonstrated superior performance on subsets of the nvBench and NL4DV benchmarks, achieving an 87\% accuracy rate, significantly outperforming existing systems like Chat2Vis and RGVisNet. This success is attributed to the iterative refinement from our multi-agent debate methodology, which rigorously ensures clarity and correctness in each script and visualization.

Nonetheless, our assessments also brought to light certain shortcomings in benchmarks such as nvBench and NL4DV, like unclear query guidelines and uneven data preparation. Developing AI-driven visualization tools requires more robust datasets, as these challenges highlight.

\subsection{Limitations}
\subsubsection{Scope of Generalization} Due to its dependence on domain-specific RoBERTa models, which necessitate unique modifications for every dataset, MASQRAD's effectiveness can be restricting. Without further training, this restricts the system's capacity to generalize to new data domains. It will be essential to use domain adaptation and advanced transfer learning techniques to address this. The flexibility and functionality of MASQRAD will be improved across a variety of domains by hybrid models that make use of pre-trained RoBERTa configurations for new domains with minimal labeled data, as well as continuous learning techniques to modify models in real-time to new data.

\subsubsection{Dynamic Schema Adaptation} MASQRAD struggles with dynamic data schemas, performing optimally only with predefined structures. This becomes problematic in environments with frequent schema changes, such as updating databases or schema-less NoSQL systems, potentially leading to inaccurate visualizations. To address this, future versions will incorporate schema detection modules capable of adapting to structural changes in real-time. These modules will employ metadata enrichment and graph-based schema inference techniques to ensure precise data representation and mapping in dynamic environments.

\subsection{Future Directions}
\subsubsection{Expanding System Capabilities}
The modularity of MASQRAD will be improved by future research, allowing for integration of reinforcement learning and APIs for more complex analysis. Because of the system's plug-and-play design, agents (like Actor AI and Critic AI) can be enhanced or swapped out for more advanced LLMs, keeping it at the forefront of technology. Healthcare, finance, and marketing are just a few of the industries that MASQRAD can adapt to by changing domain-specific models like RoBERTa or customizing prompts. Its generalization will be further evaluated through testing on various benchmarks and tasks, which will lessen the need for task-specific training.

\subsubsection{Developing Adaptive Templates}
MASQRAD will incorporate adaptive learning to refine templates dynamically based on user feedback, improving usability and precision. Schema adaptation techniques and feedback loops will help handle evolving data structures or schema-less systems. By combining continual learning with reinforcement strategies, MASQRAD will iteratively generalize across diverse query resolution tasks and industries.

\subsubsection{Overcoming Domain Dependencies}
MASQRAD's domain-specific, fine-tuned RoBERTa models, which necessitate extensive training tailored to particular datasets, are the reason for its effectiveness in NL2VIS tasks. Although accuracy is guaranteed by this specialization, the system's initial applicability to more general query resolution tasks is limited. In order to lessen these dependencies, future improvements will concentrate on combining zero-shot and few-shot learning strategies. With the help of these techniques, MASQRAD will be able to adjust more easily to a variety of tasks, such as NL2SQL or intricate decision making, requiring less retraining. This will increase its usefulness while upholding its dedication to addressing the difficulties of translating natural language into accurate visualizations, a goal shared among systems in similar area of research.

To summarize, MASQRAD represents a significant step forward in NL2VIS by refining data-driven responses through the use of a multi-agent approach. Because of its architecture's increased resilience and adaptability, automated data analysis can be implemented more widely and continue to advance. Future work will concentrate on improving MASQRAD's performance in practical situations and minimizing its dependency on preset templates, thus extending AI's potential in data analysis and visualization.

 
\bibliographystyle{IEEEtran}
\bibliography{references}

\begin{thebibliography}{10}
\providecommand{\url}[1]{#1}
\csname url@samestyle\endcsname
\providecommand{\newblock}{\relax}
\providecommand{\bibinfo}[2]{#2}
\providecommand{\BIBentrySTDinterwordspacing}{\spaceskip=0pt\relax}
\providecommand{\BIBentryALTinterwordstretchfactor}{4}
\providecommand{\BIBentryALTinterwordspacing}{\spaceskip=\fontdimen2\font plus
\BIBentryALTinterwordstretchfactor\fontdimen3\font minus \fontdimen4\font\relax}
\providecommand{\BIBforeignlanguage}[2]{{%
\expandafter\ifx\csname l@#1\endcsname\relax
\typeout{** WARNING: IEEEtran.bst: No hyphenation pattern has been}%
\typeout{** loaded for the language `#1'. Using the pattern for}%
\typeout{** the default language instead.}%
\else
\language=\csname l@#1\endcsname
\fi
#2}}
\providecommand{\BIBdecl}{\relax}
\BIBdecl

\bibitem{brown2020language}
T.~B. Brown, B.~Mann, N.~Ryder, M.~Subbiah, J.~D. Kaplan, P.~Dhariwal, A.~Neelakantan, P.~Shyam, G.~Sastry, A.~Askell \emph{et~al.}, ``Language models are few-shot learners,'' \emph{arXiv preprint arXiv:2005.14165}, 2020.

\bibitem{radford2019language}
A.~Radford, J.~Wu, R.~Child, D.~Luan, D.~Amodei, and I.~Sutskever, ``Language models are unsupervised multitask learners,'' \emph{OpenAI Blog}, vol.~1, no.~8, p.~9, 2019.

\bibitem{vaswani2017attention}
A.~Vaswani, N.~Shazeer, N.~Parmar, J.~Uszkoreit, L.~Jones, A.~N. Gomez, {\L}.~Kaiser, and I.~Polosukhin, ``Attention is all you need,'' \emph{Advances in neural information processing systems}, vol.~30, 2017.

\bibitem{he2016deep}
K.~He, X.~Zhang, S.~Ren, and J.~Sun, ``Deep residual learning for image recognition,'' in \emph{Proceedings of the IEEE conference on computer vision and pattern recognition}, 2016, pp. 770--778.

\bibitem{hochreiter1997long}
S.~Hochreiter and J.~Schmidhuber, ``Long short-term memory,'' \emph{Neural computation}, vol.~9, no.~8, pp. 1735--1780, 1997.

\bibitem{kingma2014adam}
D.~P. Kingma and J.~Ba, ``Adam: A method for stochastic optimization,'' \emph{arXiv preprint arXiv:1412.6980}, 2014.

\bibitem{lecun2015deep}
Y.~LeCun, Y.~Bengio, and G.~Hinton, ``Deep learning,'' \emph{Nature}, vol. 521, no. 7553, pp. 436--444, 2015.

\bibitem{silver2016mastering}
D.~Silver, A.~Huang, C.~J. Maddison, A.~Guez, L.~Sifre, G.~Van Den~Driessche, J.~Schrittwieser, I.~Antonoglou, V.~Panneershelvam, M.~Lanctot \emph{et~al.}, ``Mastering the game of go with deep neural networks and tree search,'' \emph{Nature}, vol. 529, no. 7587, pp. 484--489, 2016.

\bibitem{karras2019style}
T.~Karras, S.~Laine, and T.~Aila, ``A style-based generator architecture for generative adversarial networks,'' in \emph{Proceedings of the IEEE/CVF Conference on Computer Vision and Pattern Recognition}, 2019, pp. 4401--4410.

\bibitem{goodfellow2014generative}
I.~Goodfellow, J.~Pouget-Abadie, M.~Mirza, B.~Xu, D.~Warde-Farley, S.~Ozair, A.~Courville, and Y.~Bengio, ``Generative adversarial nets,'' in \emph{Advances in neural information processing systems}, 2014, pp. 2672--2680.

\bibitem{weizenbaum1966eliza}
J.~Weizenbaum, ``Eliza—a computer program for the study of natural language communication between man and machine,'' \emph{Communications of the ACM}, vol.~9, no.~1, pp. 36--45, 1966.

\bibitem{winograd1971procedures}
T.~Winograd, \emph{Procedures as a representation for data in a computer program for understanding natural language}.\hskip 1em plus 0.5em minus 0.4em\relax MIT Press, 1971.

\bibitem{ferrucci2010building}
D.~Ferrucci, E.~Brown, J.~Chu-Carroll, J.~Fan, D.~Gondek, A.~Kalyanpur, A.~Lally, J.~W. Murdock, E.~Nyberg, J.~Prager \emph{et~al.}, ``Building watson: An overview of the deepqa project,'' \emph{AI Magazine}, vol.~31, no.~3, pp. 59--79, 2010.

\bibitem{devlin2018bert}
J.~Devlin, M.-W. Chang, K.~Lee, and K.~Toutanova, ``Bert: Pre-training of deep bidirectional transformers for language understanding,'' \emph{arXiv preprint arXiv:1810.04805}, 2018.

\bibitem{chat2vis}
P.~Maddigan and T.~Susnjak, ``Chat2vis: Generating data visualizations via natural language using chatgpt, codex, and gpt-3 large language models,'' \emph{IEEE Access}, 2023.

\bibitem{woods1973progress}
W.~A. Woods, ``Progress in natural language understanding: An application to lunar geology,'' \emph{Proceedings of the National Academy of Sciences}, 1973.

\bibitem{rahman2022bert}
M.~W.~U. Rahman, S.~Shao, P.~Satam, S.~Hariri, C.~Padilla, Z.~Taylor, and C.~Nevarez, ``A bert-based deep learning approach for reputation analysis in social media,'' in \emph{2022 IEEE/ACS 19th International Conference on Computer Systems and Applications (AICCSA)}.\hskip 1em plus 0.5em minus 0.4em\relax IEEE, 2022, pp. 1--8.

\bibitem{qu2019bert}
C.~Qu, L.~Yang, M.~Qiu, W.~B. Croft, Y.~Zhang, and M.~Iyyer, ``Bert with history answer embedding for conversational question answering,'' in \emph{Proceedings of the 42nd international ACM SIGIR conference on research and development in information retrieval}, 2019, pp. 1133--1136.

\bibitem{peters2018deep}
M.~E. Peters, M.~Neumann, M.~Iyyer, M.~Gardner, C.~Clark, K.~Lee, and L.~Zettlemoyer, ``Deep contextualized word representations,'' \emph{arXiv preprint arXiv:1802.05365}, 2018.

\bibitem{raffel2019exploring}
C.~Raffel, N.~Shazeer, A.~Roberts, K.~Lee, S.~Narang, M.~Matena, Y.~Zhou, W.~Li, and P.~J. Liu, ``Exploring the limits of transfer learning with a unified text-to-text transformer,'' \emph{arXiv preprint arXiv:1910.10683}, 2019.

\bibitem{mitchell2019model}
M.~Mitchell, S.~Wu, A.~Zaldivar, P.~Barnes, L.~Vasserman, B.~Hutchinson, E.~Spitzer, I.~D. Raji, and T.~Gebru, ``Model cards for model reporting,'' in \emph{Proceedings of the Conference on Fairness, Accountability, and Transparency}, 2019, pp. 220--229.

\bibitem{bender2021dangers}
E.~M. Bender, T.~Gebru, A.~McMillan-Major, and M.~Mitchell, ``On the dangers of stochastic parrots: Can language models be too big?'' in \emph{Proceedings of the 2021 ACM Conference on Fairness, Accountability, and Transparency}.\hskip 1em plus 0.5em minus 0.4em\relax ACM, 2021, pp. 610--623.

\bibitem{tenney2019bert}
I.~Tenney, P.~Xia, B.~Chen, A.~Wang, A.~Poliak, R.~T. McCoy, N.~Kim, B.~Van~Durme, S.~R. Bowman, and D.~Das, ``What do you learn from context? probing for sentence structure in contextualized word representations,'' in \emph{International Conference on Learning Representations}, 2019.

\bibitem{yang2020joint}
Y.~Yang, M.~Richardson, M.~Surdeanu, and E.~Riloff, ``Jointly learning to extract entities and classify semantic roles,'' in \emph{Proceedings of the 58th Annual Meeting of the Association for Computational Linguistics}, 2020, pp. 6098--6109.

\bibitem{wooldridge2009introduction}
M.~Wooldridge, ``An introduction to multiagent systems,'' 2009.

\bibitem{konevcny2016federated}
J.~Kone{\v{c}}n{\`y}, H.~B. McMahan, F.~X. Yu, P.~Richt{\'a}rik, A.~T. Suresh, and D.~Bacon, ``Federated learning: Strategies for improving communication efficiency,'' in \emph{NIPS Workshop on Private Multi-Party Machine Learning}, 2016.

\bibitem{busoniu2008comprehensive}
L.~Busoniu, R.~Babuska, and B.~De~Schutter, ``A comprehensive survey of multiagent reinforcement learning,'' \emph{IEEE Transactions on Systems, Man, and Cybernetics, Part C (Applications and Reviews)}, vol.~38, no.~2, pp. 156--172, 2008.

\bibitem{kennedy1995particle}
J.~Kennedy and R.~Eberhart, ``Particle swarm optimization,'' in \emph{Proceedings of ICNN'95-International Conference on Neural Networks}, vol.~4.\hskip 1em plus 0.5em minus 0.4em\relax IEEE, 1995, pp. 1942--1948.

\bibitem{dorigo2019ant}
M.~Dorigo and T.~St{\"u}tzle, \emph{Ant colony optimization: Overview and recent advances}.\hskip 1em plus 0.5em minus 0.4em\relax Springer, 2019.

\bibitem{kraus1997negotiation}
S.~Kraus, ``Negotiation and cooperation in multi-agent environments,'' \emph{Artificial intelligence}, vol.~94, no. 1-2, pp. 79--98, 1997.

\bibitem{fatima2002multi}
S.~S. Fatima, M.~Wooldridge, and N.~R. Jennings, ``An analysis of feasible solutions for multi-agent negotiation with incomplete information,'' in \emph{Game Theory and Decision Theory in Agent-Based Systems}.\hskip 1em plus 0.5em minus 0.4em\relax Springer, 2002, pp. 99--124.

\bibitem{konda2000actor}
V.~R. Konda and J.~N. Tsitsiklis, ``Actor-critic algorithms,'' \emph{SIAM Journal on Control and Optimization}, vol.~42, no.~4, pp. 1143--1166, 2000.

\bibitem{liu2019roberta}
Y.~Liu, M.~Ott, N.~Goyal, J.~Du, M.~Joshi, D.~Chen, O.~Levy, M.~Lewis, L.~Zettlemoyer, and V.~Stoyanov, ``Roberta: A robustly optimized bert pretraining approach,'' \emph{arXiv preprint arXiv:1907.11692}, 2019.

\bibitem{xu2023understanding}
W.~Xu, S.~Agrawal, E.~Briakou, M.~J. Martindale, and M.~Carpuat, ``Understanding and detecting hallucinations in neural machine translation via model introspection,'' \emph{Transactions of the Association for Computational Linguistics}, vol.~11, pp. 546--564, 2023.

\bibitem{smit2023we}
A.~Smit, P.~Duckworth, N.~Grinsztajn, K.-a. Tessera, T.~D. Barrett, and A.~Pretorius, ``Are we going mad? benchmarking multi-agent debate between language models for medical q\&a,'' \emph{arXiv preprint arXiv:2311.17371}, 2023.

\bibitem{bai2022constitutional}
Y.~Bai, S.~Kadavath, S.~Kundu, A.~Askell, J.~Kernion, A.~Jones, A.~Chen, A.~Goldie, A.~Mirhoseini, C.~McKinnon \emph{et~al.}, ``Constitutional ai: Harmlessness from ai feedback,'' \emph{arXiv preprint arXiv:2212.08073}, 2022.

\bibitem{claude2023}
Anthropic, ``Introducing the next generation of claude,'' \url{https://www.anthropic.com}, 2023, accessed: 2023-09-30.

\bibitem{openai2024gpt4o}
\BIBentryALTinterwordspacing
OpenAI, ``Hello gpt-4o,'' 2024, accessed: 2024-07-09. [Online]. Available: \url{https://openai.com/index/hello-gpt-4o/}
\BIBentrySTDinterwordspacing

\bibitem{anthropic2023claude35}
\BIBentryALTinterwordspacing
Anthropic, ``Claude 3.5 sonnet model card addendum,'' 2023, accessed: 2024-07-09. [Online]. Available: \url{https://www-cdn.anthropic.com/fed9cc193a14b84131812372d8d5857f8f304c52/Model_Card_Claude_3_Addendum.pdf}
\BIBentrySTDinterwordspacing

\bibitem{liu2021pretrain}
P.~Liu, W.~Yuan, J.~Fu, Z.~Jiang, H.~Hayashi, and G.~Neubig, ``Pre-train, prompt, and predict: A systematic survey of prompting methods in natural language processing,'' \emph{arXiv preprint arXiv:2107.13586}, 2021.

\bibitem{gao2021making}
T.~Gao, A.~Fisch, and D.~Chen, ``Making pre-trained language models better few-shot learners,'' in \emph{Proceedings of the 59th Annual Meeting of the Association for Computational Linguistics (ACL-IJCNLP)}, 2021, pp. 3816--3830.

\bibitem{lester2021power}
B.~Lester, R.~Al-Rfou, and N.~Constant, ``The power of scale for parameter-efficient prompt tuning,'' in \emph{Proceedings of the 2021 Conference on Empirical Methods in Natural Language Processing}, 2021, pp. 3045--3059.

\bibitem{luo2021nvbench}
Y.~Luo, J.~Tang, and G.~Li, ``nvbench: A large-scale synthesized dataset for cross-domain natural language to visualization task,'' \emph{arXiv preprint arXiv:2112.12926}, 2021.

\bibitem{narechania2020nl4dv}
A.~Narechania, A.~Srinivasan, and J.~Stasko, ``Nl4dv: A toolkit for generating analytic specifications for data visualization from natural language queries,'' \emph{IEEE Transactions on Visualization and Computer Graphics}, vol.~27, no.~2, pp. 369--379, 2020.

\bibitem{podo2024vi}
L.~Podo, M.~Ishmal, and M.~Angelini, ``Vi (e) va llm! a conceptual stack for evaluating and interpreting generative ai-based visualizations,'' \emph{arXiv preprint arXiv:2402.02167}, 2024.

\bibitem{luo2021synthesizing}
Y.~Luo, N.~Tang, G.~Li, C.~Chai, W.~Li, and X.~Qin, ``Synthesizing natural language to visualization (nl2vis) benchmarks from nl2sql benchmarks,'' in \emph{Proceedings of the 2021 International Conference on Management of Data}, 2021, pp. 1235--1247.

\bibitem{luo2021natural}
Y.~Luo, N.~Tang, G.~Li, J.~Tang, C.~Chai, and X.~Qin, ``Natural language to visualization by neural machine translation,'' \emph{IEEE Transactions on Visualization and Computer Graphics}, vol.~28, no.~1, pp. 217--226, 2021.

\bibitem{song2022rgvisnet}
Y.~Song, X.~Zhao, R.~C.-W. Wong, and D.~Jiang, ``Rgvisnet: A hybrid retrieval-generation neural framework towards automatic data visualization generation,'' in \emph{Proceedings of the 28th ACM SIGKDD Conference on Knowledge Discovery and Data Mining}, 2022, pp. 1646--1655.

\bibitem{li2024visualization}
G.~Li, X.~Wang, G.~Aodeng, S.~Zheng, Y.~Zhang, C.~Ou, S.~Wang, and C.~H. Liu, ``Visualization generation with large language models: An evaluation,'' \emph{arXiv preprint arXiv:2401.11255}, 2024.

\end{thebibliography}

\begin{IEEEbiography}[{\includegraphics[width=1.05in,height=1.3125in,clip,keepaspectratio]{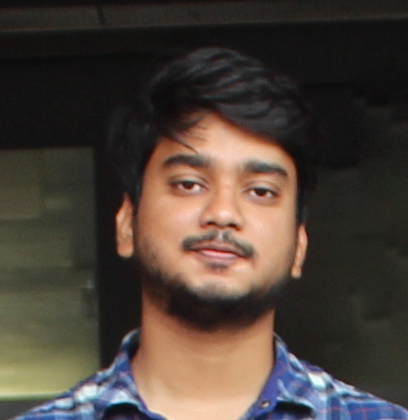}}]{Mohammad Wali Ur Rahman}
graduated with a Master of Science degree in Electrical and Computer Engineering from the University of Arizona in 2023 and is currently pursuing his Ph.D. in the same field at the same institution. He has been engaged as a Research Assistant at the Autonomic Computing Lab, a branch of the NSF-CAC (NSF Center for Autonomic Computing). His academic and research pursuits are centered around Text Mining \& Analysis, Natural Language Processing, Machine Learning, Neural Networks, Artificial Intelligence and cybersecurity.
\end{IEEEbiography}

\begin{IEEEbiography}[{\includegraphics[width=1.05in,height=1.3125in,clip,keepaspectratio]{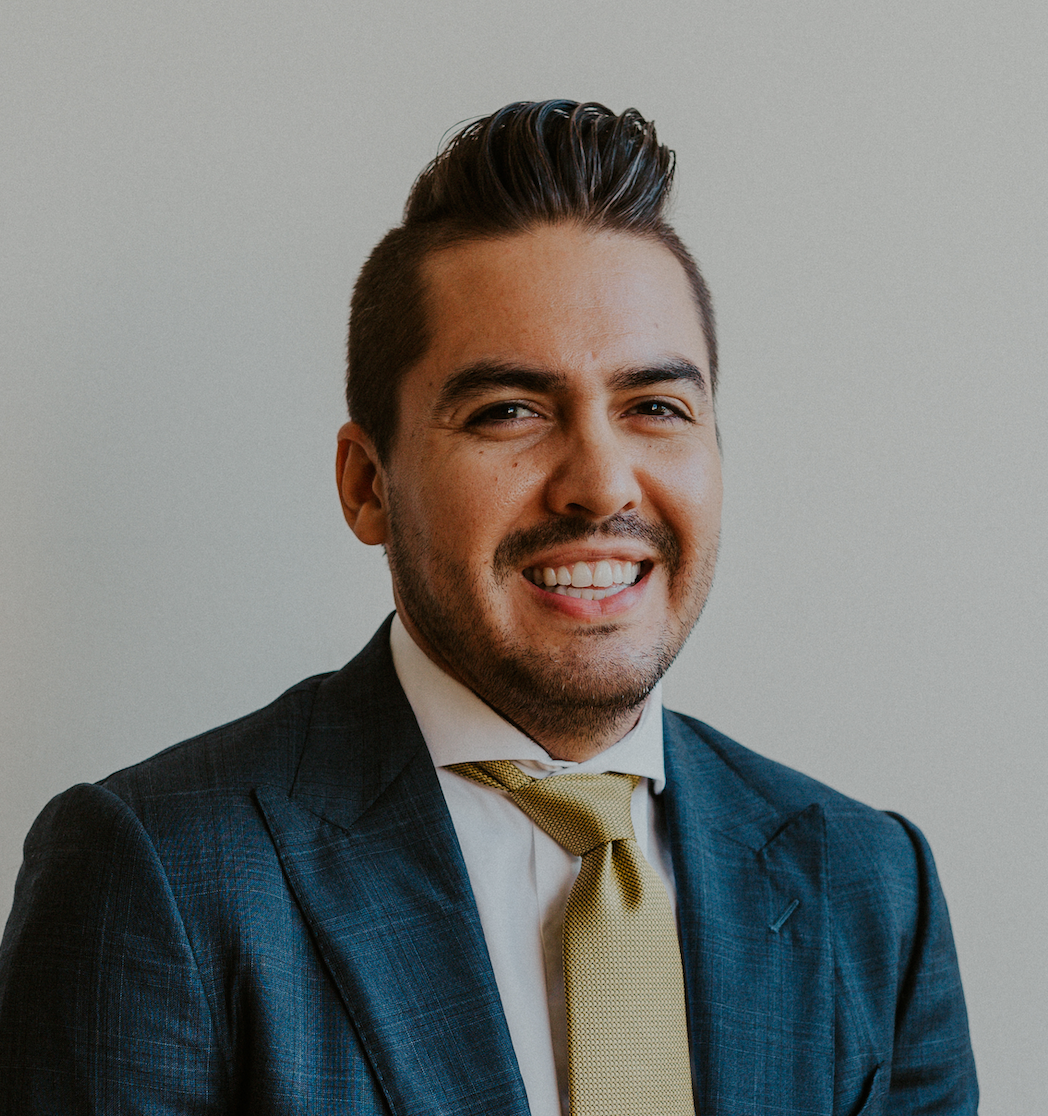}}]{Ric Nevarez}
is the Chief Technology Officer at Trust Web and holds a Bachelor of Science degree in Nutritional Biochemistry from Brigham Young University. With over 14 years of experience in the artificial intelligence and natural language processing industry, Ric Nevarez has been a strong advocate for sustainable and ethical AI practices. As an experienced product manager and entrepreneur, he has co-founded several successful AI startups, securing substantial funding and developing innovative AI products. 
\end{IEEEbiography}

\begin{IEEEbiography}[{\includegraphics[width=1.05in,height=1.3125in,clip,keepaspectratio]{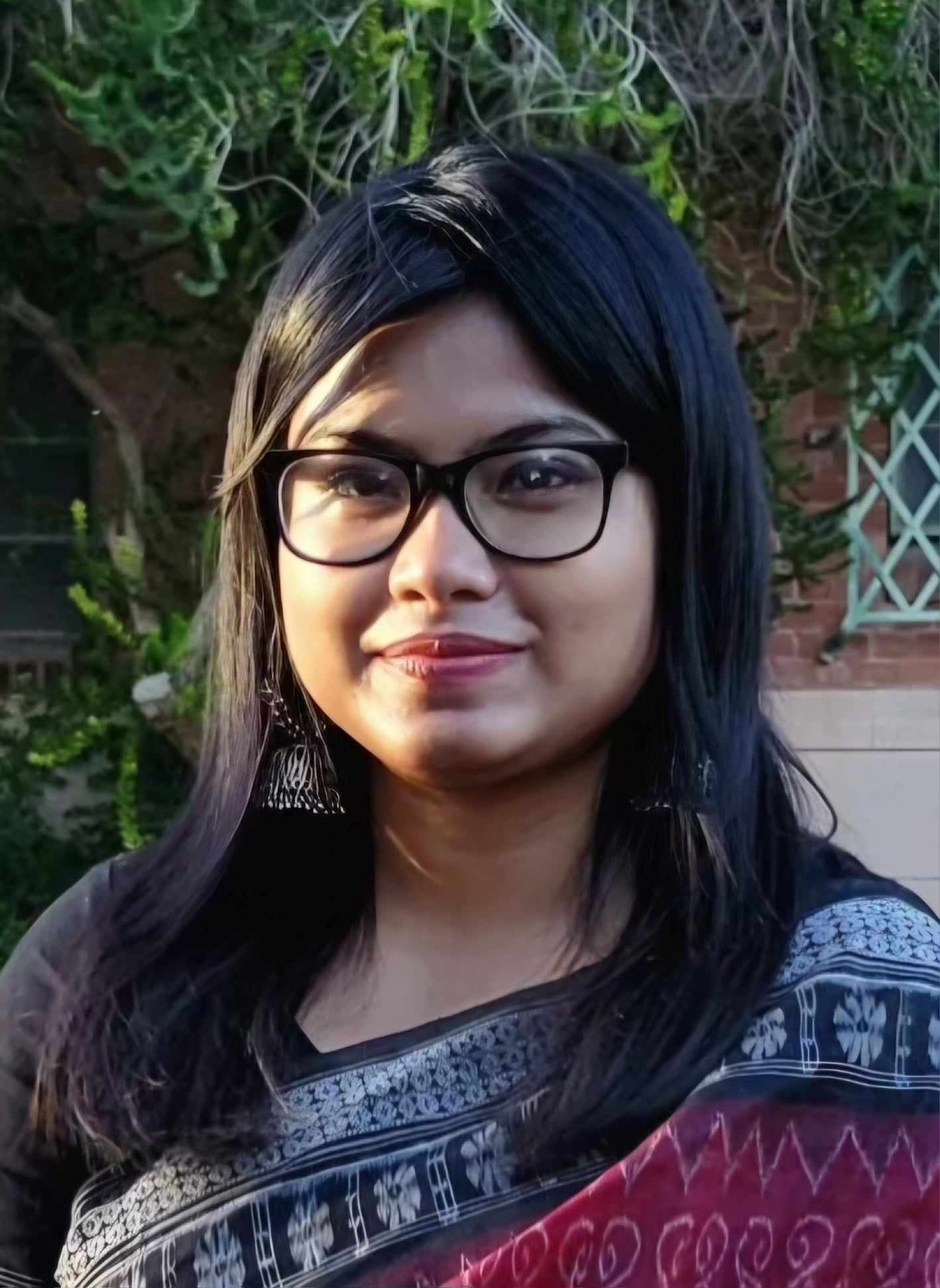}}]{Lamia Tasnim Mim}
 is a graduate of New Mexico State University, where she earned her Master's in Computer Science (2023). She specializes in Artificial Intelligence, Machine Learning, and Natural Language Processing, with a focus on developing robust and scalable deep learning models. Currently, Lamia serves as a Machine Learning Engineer at Avirtek, Inc., where she develops streaming data pipelines, deploys ML models, and enhances data recovery accuracy through advanced techniques. 
\end{IEEEbiography}

\begin{IEEEbiography}[{\includegraphics[width=1.05in,height=1.3125in,clip,keepaspectratio]{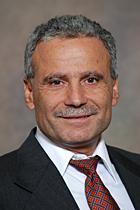}}]{Dr Salim Hariri}
(Senior Member, IEEE) received an
M.Sc. degree from Ohio State University in 1982, and a Ph.D. degree in Computer Engineering from the University of Southern California in 1986. He is a Professor in the Department of Electrical and Computer Engineering, the University of Arizona, and the Director of the NSF Center for Cloud and Autonomic Computing (NSF-CAC). His research focuses on autonomic computing, artificial intelligence, machine learning, cybersecurity, cyber resilience, and cloud security.
\end{IEEEbiography}
\vspace{-10mm} 

\end{document}